%% file: main.tex
\documentclass[sigconf, screen]{acmart}

\renewcommand\footnotetextcopyrightpermission[1]{} % removes footnote with conference information in first column
\setcopyright{none}
\usepackage{tikz}
\usepackage{amsmath}
\usepackage{xspace}
\usepackage[ruled,vlined,linesnumbered]{algorithm2e}
\usepackage{graphicx}
\usepackage{caption}
\usepackage{subcaption}
\usepackage{wasysym}
\usepackage{multirow}
\usepackage{array}
\usepackage{enumitem}
\PassOptionsToPackage{table,usenames,dvipsnames}{xcolor}

\usepackage{balance}
\usepackage{dsfont}

\usepackage[skip=2pt]{caption}
\usepackage[skip=0pt]{subcaption}

\newcommand{\TODO}[1]{{\color{red}TODO: #1}}
\newtheorem{definition}{Definition}

\newcommand{\Hammer}{\emph{Speed-ANN}\xspace}
\newcommand{\SeqFullName}{Best-First Search\xspace}
\newcommand{\SeqShortName}{\emph{BFiS}\xspace}
\newcommand{\TopM}{\Hammer}
\newcommand{\TopMShortName}{\Hammer}

\newcommand{\PunchStarter}[1]{\noindent\textbf{#1}\xspace}

\definecolor{SeafoamGreen}{RGB}{159, 226, 191}
\definecolor{Peach}{RGB}{255, 203, 164}

\acmConference[Arxiv '22]{Arxiv}{2022}{}
\acmBooktitle{Arxiv, 2022}
\begin{document}
\title{
    %\Hammer: Efficient Nearest Neighbor Search via Intra-Query Parallelism and Cache Friendly Indexing
%    iQANS: Low-Latency and High-Accuracy Billion-Scale Nearest Neighbor Search via Intra-Query Parallelism
    Speed-ANN: Low-Latency and High-Accuracy Nearest Neighbor Search via Intra-Query Parallelism
}
%%
%% The "author" command and its associated commands are used to define the authors and their affiliations.
\author{Zhen Peng}
\affiliation{%
    \institution{College of William \& Mary}
    % \streetaddress{P.O. Box 1212}
    \city{Williamsburg}
%    \state{Ireland}
    \country{VA USA}
    \postcode{43017-6221}
}
\email{zpeng01@wm.edu}

\author{Minjia Zhang}
% \orcid{0000-0002-1825-0097}
\affiliation{%
    \institution{Microsoft AI and Research}
    % \streetaddress{1 Th{\o}rv{\"a}ld Circle}
    \city{Bellevue}
    \country{WA USA}
}
\email{minjiaz@microsoft.com}

\author{Kai Li}
% \orcid{0000-0001-5109-3700}
\affiliation{%
    \institution{Kent State University}
    \city{Kent}
    \country{OH USA}
}
\email{kli17@kent.edu }

\author{Ruoming Jin}
\affiliation{%
    \institution{Kent State University}
    \city{Kent}
    \country{OH USA}
}
\email{rjin1@kent.edu}
% \email{myprivate@email.com}
% \email{second@affiliation.mail}

% \author{Wang Xiu Ying}
% \author{Zhe Zuo}
% \affiliation{%
%     \institution{East China Normal University}
%     \city{Shanghai}
%     \country{China}
% }
% \email{firstname.lastname@ecnu.edu.cn}

\author{Bin Ren}
\affiliation{%
    \institution{College of William \& Mary}
    \city{Williamsburg}
    \country{VA USA}
}
% \affiliation{%
%     \institution{Donald's Second Affiliation}
%     \city{City}
%     \country{country}
% }
\email{bren@cs.wm.edu}

%%%% body
%-------------------------------------------------------------------------------
\begin{abstract}
%-------------------------------------------------------------------------------
%Your abstract text goes here. Just a few facts. Whet our appetites.
%Not more than 200 words, if possible, and preferably closer to 150.

%\textcolor{red}{TODO: placeholder}
%\bin{TODO: uniform all terms}

Nearest Neighbor Search (NNS) has recently drawn a rapid increase of interest due to its core role in managing high-dimensional vector data in data science and AI applications. The interest is fueled by the success of neural embedding, where deep learning models transform unstructured data into semantically correlated feature vectors for data analysis, e.g., recommend popular items. Among several categories of methods for fast NNS, similarity graph is one of the most successful algorithmic trends. Several of the most popular and top-performing similarity graphs, such as NSG and HNSW, at their core employ best-first traversal along the underlying graph indices to search near neighbors. Maximizing the performance of the search is essential for many tasks, especially at the large-scale and high-recall regime.  
In this work, we provide an in-depth examination of the challenges of the state-of-the-art similarity search algorithms, revealing its challenges in leveraging multi-core processors to speed up the search efficiency. We also exploit whether similarity graph search is robust to deviation from maintaining strict order by allowing multiple walkers to simultaneously advance the search frontier. Based on our insights, we propose \Hammer, a parallel similarity search algorithm that exploits hidden intra-query parallelism and memory hierarchy that allows similarity search to take advantage of multiple CPU cores to significantly accelerate search speed while achieving high accuracy.

We evaluate \Hammer on a wide range of datasets, ranging from million to billion data points, and show that it reduces query latency by $2.1\times$, $5.2\times$, and $13\times$ on average than NSG and $2.1\times$, $6.7\times$, and $17.8\times$ on average than HNSW at 0.9, 0.99, and 0.999 recall target, respectively.
More interesting, our approach achieves super-linear speedups in some cases using 32 threads, achieving up to 37.7 times and 76.6 times faster to obtain the same accuracy than two state-of-the-art graph-based nearest neighbor search methods NSG and HNSW, respectively. Finally, with multicore support, we show that our approach offers faster search latency than highly-optimized GPU implementation and provides good scalability as the increase of the number of hardware resources (e.g., CPU cores) and graph sizes, offering up to $16.0\times$ speedup on two {billion-scale} datasets.  
\end{abstract}

% restraining the overall distance computation and avoiding high synchronization overhead. 

\maketitle

\input{text/intro}

\input{text/background}
\input{text/overview}

% \input{text/design}
\input{text/designv2}
\input{text/eval}
\input{text/related}
\input{text/conclusion}

%\bibliographystyle{plain}
\bibliographystyle{ACM-Reference-Format}
\bibliography{bib/ref.bib,bib/refnew.bib}

\end{document}

%% file: text/intro.tex
\section{Introduction}\label{sec:intro}

%\textcolor{red}{TODO: placeholder}

% 1. ML + Vector Database kernel/fundamental operation -- Similarly search;

% 2. Unique graph problem, different from BSP. {\em Target-oriented} Any bound/math

% 3. Parallelization. Lock-step to speculative with computation limitation.

% 4. Result

% Intra-query; query latency.
% example (100ms for high recall cannot meet requirement)

Nearest neighbor search (NNS) is a fundamental building block for many applications within machine learning systems and database management systems, such as recommendation systems~\cite{das2007google}, large-scale image search and information retrieval~\cite{kulis2009kernelized,lv2004image,philbin2007object}, 
entity resolution~\cite{hoffart2012kore}, and sequence matching~\cite{berlin2015assembling}. NNS has recently become the focus of intense research activity, due to its core role in semantic-based search of unstructured data such as images, texts, video, speech using neural embedding models. In semantic-based search, a neural embedding model transfers objects into \emph{embeddings} in $\mathds{R}^d$, where $d$ often ranges from 100 to 1000 and N ranges from millions to billions. The task then is to find the $K$ nearest embeddings for a given query. For example, major e-commerce players such as Amazon~\cite{amazon-search} and Alibaba~\cite{alibaba-search} build semantic search engines, which embed product catalog and the search query into the same high-dimensional space and then recommends products whose embeddings that are closest to the embedded search query; Youtube~\cite{youtube-embed} embeds videos to vectors for video recommendation; Web-scale search engines embed text (e.g., word2vec~\cite{word2vec}, doc2vec~\cite{doc2vec}) and images (e.g., VGG~\cite{vgg}) for text/image retrieval~\cite{sptag,rankbrain}. We expect applications built on top of the embedding-based search to continue growing in the future, due to the success and continual advancement of neural embedding techniques that can effectively capture the semantic relations of objects. We also expect the objects to embed will grow rapidly, due to ubiquitous data collections, e.g., through phones and IoT devices.

Since the search occurs for every query, the \emph{latency} and the \emph{accuracy} (\emph{recall}) of the search engine critically depend on the ability to perform fast near neighbor search in the high-recall range. Various solutions for approximate nearest neighbor search (ANNS) have been proposed, including hashing-based methods\cite{indyk1998approximate,datar2004locality,andoni2006near,andoni2015practical}, quantization-based methods~\cite{jegou2008hamming,ge2013optimized,wu2017multiscale,wei2020analyticdb}, tree-based methods~\cite{silpa2008optimised,beckmann1990r,wang2020deltapq}, and graph-based methods~\cite{malkov2014approximate,wu2014fast,fu2019fast}. Among them, graph-based algorithms have emerged as a remarkably effective class of methods for high-dimensional ANNS, outperforming other approaches for very high recalls on a wide range of datasets~\cite{ann-benchmark}. As a result, these graph-based algorithms have been integrated with many large-scale production systems~\cite{fu2019fast,malkov2020efficient}, where optimizations for fast search and high recall are the focus of a highly active research area and have a clear practical impact.

To provide scalability, existing ANN search libraries often resort to coarse-grained inter-query parallelism, by dispatching each query to a core or even across different machines such that multiple queries can be processed simultaneously~\cite{fu2019fast,bashyam2020fast}. Although inter-query parallelism obtains impressive throughput improvements, it does not help reduce query latency. In particular, online applications often process each query upon its arrival and have stringent latency service level agreements (e.g., a few milliseconds). As the size of datasets grows rapidly, the increased latency of current graph-based ANN algorithms has been restraining ANN-based search engines from growing to large-scale datasets, especially for high-recall regimes.
To provide relevant results with consistently low latency, in this work, we investigate the possibility of intra-query parallelism on individual nodes to meet latency goals. 

Although graph-based ANN consists of primarily graph operations, simply dividing the work of graph traversal into multiple threads is insufficient for supporting efficient ANN search, as it cannot efficiently leverage the underline multi-core processors due to complex interactions between graph operations and the hardware threads and memory hierarchy. In our studies, the intra-query parallelism may sometimes hurt search efficiency, because the communication and synchronization overhead increases as we increase the number of cores, making it especially harder to achieve high efficiency.

In this work, we provide an in-depth examination of the graph-based ANN algorithms with intra-query parallelism. Through a series of experiments, we have identified that an intrinsic challenge of the graph search process lies in its long convergence step --- existing \emph{best-first search} leads to long convergence steps and introduces heavy control dependencies that limit the upper bounds on speedup by using more cores, as predicted by Amdahl's Law. In our study, we find that, by enlarging the \SeqFullName to \TopM, the search process can converge in much fewer iterations, suggesting that the search process can achieve better overall performance by running individual queries with more hardware resources. However, exposing the \emph{path-wise} parallelism also changes the search dynamics of queries, leading to additional challenges that may adversely affect search efficiency, which resides in the aspects of redundant computations, memory-bandwidth under-utilization, high synchronization overhead, and irregular accesses caused poor data locality. 

Based on the insights from our analysis, we present \Hammer, a similarity search algorithm that combines a set of optimizations to address these challenges. \Hammer introduces three tailored optimizations to provide improved performance for graph-based ANN search. 
First, \Hammer uses \emph{parallel neighbor expansion} to divide the search workload to multiple workers in coarse-grained parallelism. Among it, every worker performs its private best-first search in an asynchronous manner to avoid heavy global communication.
Second, \Hammer employs a \emph{staged search} scheme, which reduces redundant computations caused by over-expansion during a parallel search. 
Third, \Hammer is characterized by \emph{redundant-expansion aware synchronization} to lazily synchronize among workers while still providing fast search speed high recall.
Finally, \Hammer provides additional optimizations such as loosely synchronized visiting maps and a cache-friendly neighbor grouping mechanism to improve cache locality during parallel search.
In summary, this paper makes the following contributions:
\begin{enumerate}[leftmargin=0.24in,topsep=0pt]
\item provides the first comprehensive experimental analysis of intra-query parallelism for ANN search on multi-core architecture and identifies several major bottlenecks to speedup graph-based approximate nearest neighbors in high recall regime; 
\item studies how the characteristics of a query vary as the search moves forward from multiple aspects, e.g., by increasing the edge-wise parallelism degree and the dynamics in search queue update positions, which reveals the opportunities and challenges it brings;
\item introduces a search algorithm named \Hammer with novel optimizations such as \emph{staged parallel neighbor expansion} and \emph{redundant-expansion aware synchronization} that allow parallel search on graph-based ANN to achieve significantly lower latency with high recall on different multi-core hardwares. 
\item conducts thorough evaluation on a wide range of real-world datasets ranging from million to {\bf billion} data points to show that \Hammer speeds up the search by $1.3\times$--$76.6\times$ compared to highly optimized state-of-the-art CPU-based search algorithms NSG~\cite{fu2019fast} and HNSW~\cite{malkov2020efficient}. \Hammer sometimes achieves super-linear speedups in the high recall regime as the number of threads increases, obtaining up to $37.7\times$ speedup over NSG and up to $76.6\times$ speedup over HNSW when using 32 threads. \Hammer also outperforms a state-of-the-art GPU implementation and provides good scalability.
\end{enumerate}

%%%% We evaluate {\bf NAME} on XX datasets and compare it with XX state-of-the-art approaches, XX and XX, achieving XX to XX $\times$ speedup.

%, offering up to $16.0\times$ speedup on two {\bf billion-scale} datasets.

%% file: text/background.tex
\section{Preliminaries}\label{sec:background}

\subsection{Approximate Nearest Neighbors}\label{subsec:problem_setting}

% \subsection{Nearest Neighbor Search Problem}
% Define the Nearest Neighbor Search and Approximate Nearest Neighbor Search.
%\textcolor{red}{TODO: cut some content}

The nearest neighbor search problem in high-dimensional space is fundamental in various applications of information retrieval and database management. In this paper, the Euclidean space under the $l_2$ norm is denoted by $E^d$. The closeness of any two points $p_1$ and $p_2$ is defined by the $l_2$ distance $\delta (p_1, p_2)$ between them~\cite{fu2019fast}.
The \emph{Nearest Neighbor Search (NNS)} can be defined as follows~\cite{gionis1999similarity}:

\begin{definition}[\textbf{Nearest Neighbor Search}]\label{def:NNS}
    Given a finite point set $P$ of $n$ points in the space $E^d$, preprocess $P$ so as to answer a given query point $q$ by finding the closest point $p \in P$.
\end{definition}

Please note that the query point $q$ is not in the point set $P$, i.e. $q \notin P$. The above definition generalizes naturally to the \emph{$K$ Nearest Neighbor Search (K-NNS)} where we want to find $K > 1$ points in the database that are closest to the query point.
A na\"ive solution is to linearly iterate all points in the dataset and evaluate their distance to the query. It is computationally demanding and only suitable for small datasets or queries without a time limit of response. 
Therefore, it is practical to relax the condition of the exact search by allowing some extent of approximation.
The \emph{Approximate Nearest Neighbor Search (ANNS)} problem can be defined as follows~\cite{gionis1999similarity}:

\begin{definition}[\textbf{$\epsilon$-Nearest Neighbor Search}]\label{def:ANNS}
    Given a finite point set $P$ of $n$ points in the space $E^d$, preprocess $P$ so as to answer a given query point $q$ by finding a point $p \in P$ such that $\delta (p, q) \leq (1 + \epsilon) \delta (r, q)$ where $r$ is the closest point to $q$ in $P$.
\end{definition}

Similarly, this definition can generalize to the \emph{Approximate $K$ Nearest Neighbor Search (AKNNS)} where we wish to find $K > 1$ points $p_1, \ldots, p_K$ such that $\forall i = 1, \ldots,K, \delta (p_i, q) \leq (1 + \epsilon) \delta (r_i, q)$ where $r_i$ is the $i$th closest point to $q$.

In practice, determining the exact value of $\epsilon$ requires some hard efforts. Instead, we use \emph{recall} as the metric to evaluate the quality of the approximation. A high \emph{recall} implies a small $\epsilon$, thus a good quality of the approximation. 
It is defined as the value of the recall. 
Suppose the approximate points set found for a given query $q$ is $R'$, and the true $K$ nearest neighbor set of $q$ is $R$, the \emph{recall} is
%\minjia{@Zhen. Change "accuracy" to "recall".}
defined as follows~\cite{fu2016efanna}:
\begin{align} \label{formula:recall}
    Recall(R') = \frac{\left | R' \cap R \right |}{\left | R' \right |} = \frac{\left | R' \cap R \right |}{K}
\end{align}

For a particular recall target, i.g. $0.990$ or $0.995$, our goal is to make the query latency as short as possible.

% Define the recall.

\subsection{Graph-based ANN Search}

\begin{figure}
    \centering
    \begin{subfigure}[t]{0.15\textwidth}
        \centering
        \includegraphics[width=\textwidth]{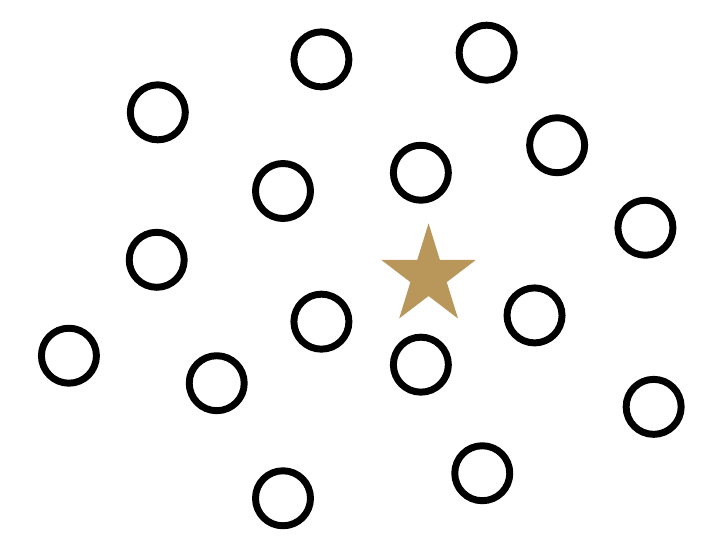}
        \caption{Data points and a query point.}
        \label{subfig:dataset_and_query}
    \end{subfigure}
    \hfill
    \begin{subfigure}[t]{0.15\textwidth}
        \centering
        \includegraphics[width=\textwidth]{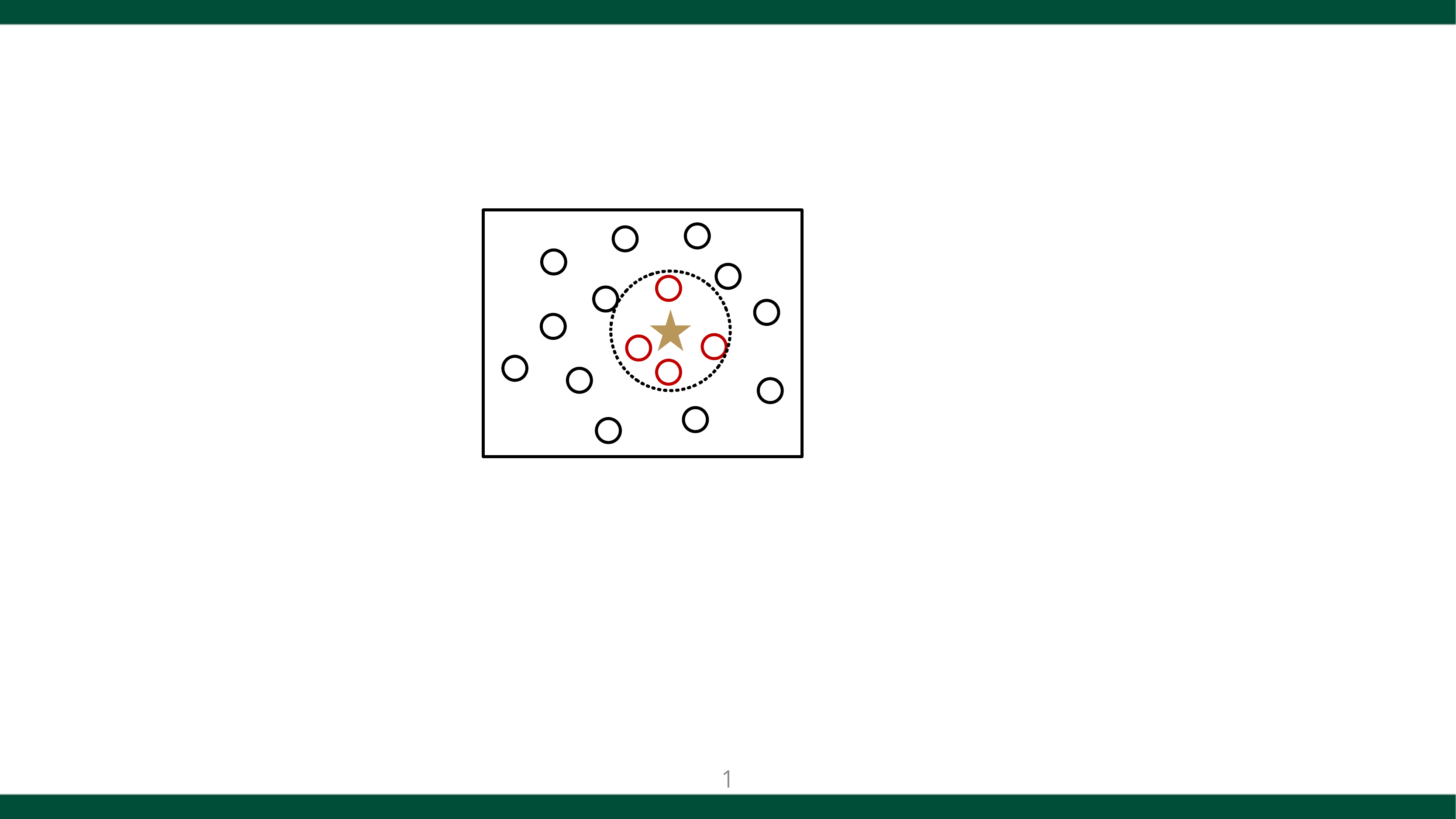}
        \caption{Nearest neighbors of the query.}
        \label{subfig:nearest_neighbors}
    \end{subfigure}
    \hfill
    \begin{subfigure}[t]{0.15\textwidth}
        \centering
        \includegraphics[width=\textwidth]{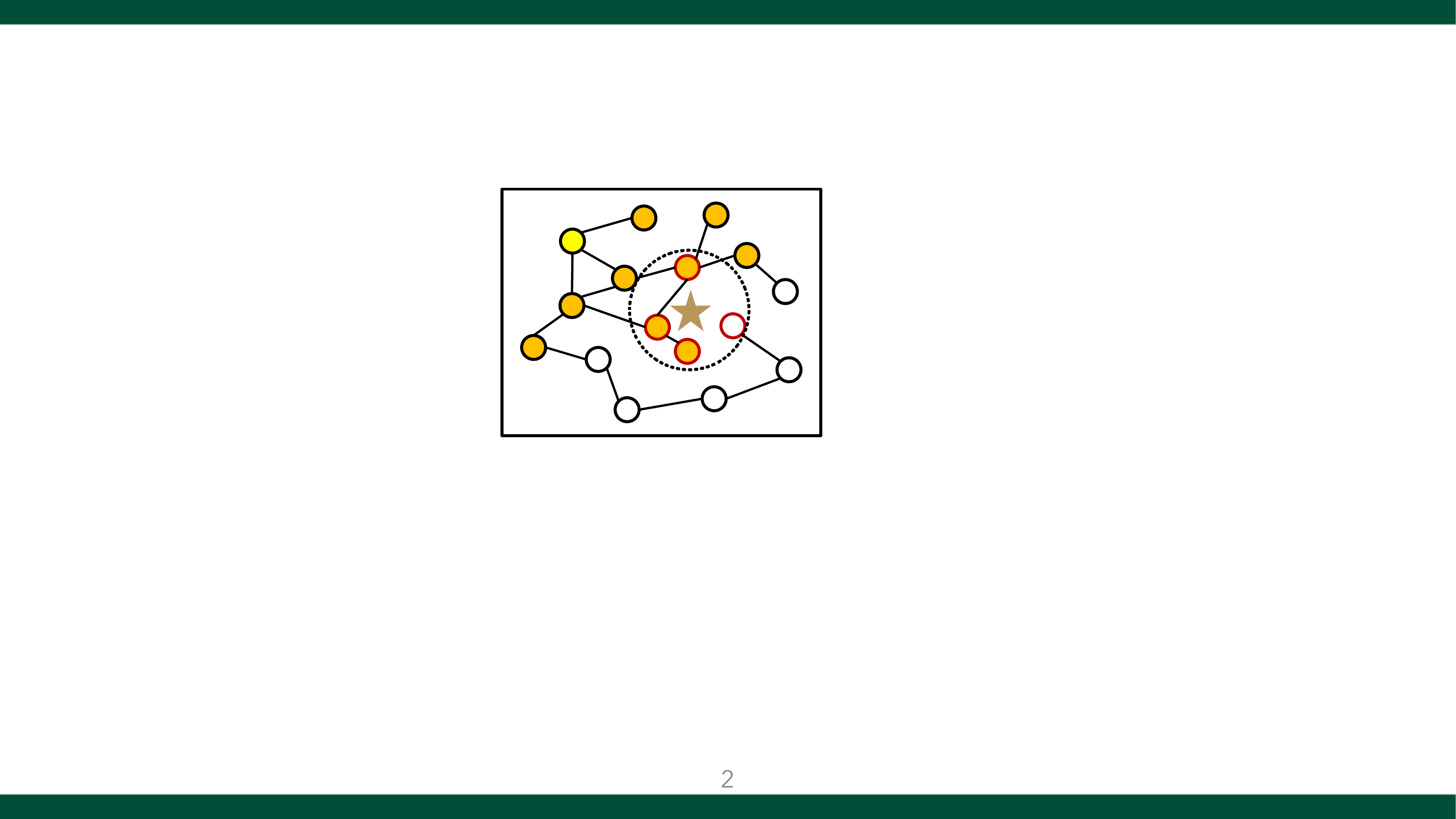}
        \caption{A graph index and search procedure.}
        \label{subfig:graph_index}
    \end{subfigure}
    \caption{An example of graph-based ANNS. \textmd{Circles are data points. The golden star is query target (not in dataset). Four red circles are its nearest neighbors. Graph-based ANNS builds a graph index on the dataset in~\ref{subfig:graph_index}. The yellow circle is the starting point. Orange circles are visited vertices during the search via Algorithm~\ref{algo:seq_greedy_search}.}}
    \label{fig:graph_based_anns}
    \vspace{1em}
\end{figure}

% \begin{figure}
%     \centering
%     \includegraphics[width=0.45\textwidth]{figures/fig_seq_search}
%     \caption{The search procedure of \SeqFullName. 
%         \textmd{The procedure is a sequence of states of the priority queue which stores the nearest vertices so far. Beginning with the starting point, the queue is updated by greedily expanding the first unchecked candidate and inserting its neighbors as new candidates. It stops when candidates in the queue are all checked and no more updates, and then returns nearest neighbors found.}}
%     \label{fig:fig_seq_search}
%     \minjia{This figure is referenced very lightly in the text (Section 3.1 and Challenge 1 in Section 3.2) and seems to be duplicated with Algorithm 1. My suggestion is to remove it if we are short for space.}
% \end{figure}

% Introduce graph-based methods, especially NSG~\cite{fu2019fast}.

%As a fundamental machine learning operation, 
Various ANNS solutions have been proposed over decades, e.g., the ones based on trees~\cite{arora2018hd}, hashing~\cite{huang2015query}, quantization~\cite{andre2015cache}, and graphs~\cite{malkov2014approximate,wu2014fast,fu2019fast}. Recently, many experimental results~\cite{malkov2014approximate,fu2019fast} show that graph-based approaches usually outperform others, 
resulting in the best execution performance and recall. 
That is because graph-based approaches can better express the neighbor relationship, allowing to check much fewer points in neighbor-subspaces.

%A graph is an abstract structure which consists of a set of vertices and edges that link some pairs of the vertices.
%For the ANNS problem defined above, a graph can be used as the index structure for navigating where 
Graph-based ANNS relies on a graph structure as its index, in which
a vertex represents a data point in the data set and an edge links two points as shown in Figure~\ref{fig:graph_based_anns}. A vertex $p_2$ is called a \emph{neighbor} of a vertex $p_1$ if and only if there is an edge from $p_1$ to $p_2$.
Many prior efforts focus on constructing optimal graphs for efficient ANNS~\cite{fu2019fast,malkov2020efficient}---which is not the focus of this work. In contrast, this work is based on the state-of-the-art graph construction approach~\cite{fu2019fast}, and aims to parallelize ANNS itself with a thorough study of its bottleneck and a set of advanced techniques addressing these bottlenecks.  

Given the graph-based index built ready, \SeqFullName algorithm is widely used by many graph-based methods for searching nearest neighbors~\cite{dearholt1988monotonic,arya1993approximate,hajebi2011fast,jin2014fast,malkov2014approximate,malkov2020efficient,harwood2016fanng}. 
Given a query point $Q$ and a starting point $P$, the algorithm is to search for $K$ nearest neighbors to $Q$.
In the first search step, it visits $P$'s neighbors and computes their distance to $Q$ respectively to choose the closest vertex or candidate, and the next search step starts from the chosen candidate from the last step.
All visited vertices are recorded and kept in order according to their distance to $Q$. 
The search step stops when the first $K$ visited vertices do not change anymore, which are the final $K$ nearest neighbors. 
The time spent to find the $K$ nearest neighbors is the query's \emph{latency}.

\iffalse%%%%%%%%%%%

The \SeqFullName (\SeqShortName) is the baseline ANNS algorithm used in NSG~\cite{fu2019fast} and other graph based index methods. 
%Algorithm~\ref{algo:seq_greedy_search} introduces the baseline sequential algorithm that is used by ~\cite{fu2019fast} to perform ANNS on the constructed graph.
%As shown in Algorithm~\ref{algo:seq_greedy_search}, its basic idea is to start from one point, and navigate the graph in a greedy manner, during which it records those current closest candidate vertices in a priority queue until no new (unchecked) vertex can add to this queue any more. 
%In the priority queue, all vertices are sorted according to their distance to the query.
As shown in Algorithm~\ref{algo:seq_greedy_search}, its basic idea is to start from one point, and navigate the graph in a greedy manner, during which it records those current closest candidate vertices in a priority queue where all candidates are sorted regarding their distance to the query.
The whole search procedure can be regarded as a sequence of state update for the priority queue, as shown in Figure~\ref{fig:fig_seq_search}.
In every state, the best first unchecked candidate is selected out and then expanded, whose neighbors are visited and their distance to the query is computed. Those neighbors are then inserted into the queue as new unchecked candidates. The search procedure terminates or converges when no new (unchecked) vertex can update the queue.

\input{text/algo_seq_greedy_search}

\fi%%%%%%%%%%%%%%%%%%%%%%

%% file: text/overview.tex
\section{Complexities in Graph-based ANN Search for Optimizations}\label{sec:overview}

%\minjia{@Zhen, the header has "July 2017". Needs to make it up-to-date, say 2021.}

\subsection{Overview of Graph-based ANN Search}

%\textcolor{red}{TODO: formal definition}
%\textcolor{red}{TODO: new computation abstraction in Figure 2}

The search procedure in existing similarity graph algorithms, such as NSG~\cite{fu2019fast} and HNSW~\cite{malkov2020efficient}, is a \emph{best-first traversal} that starts at a chosen (e.g., medoid o random) point and walks along the edges of the graph while getting closer to the nearest neighbors at each step until it converges to a local minimum.  Algorithm~\ref{algo:seq_greedy_search} shows its basic idea. In a similarity graph, nodes represent entities in a problem domain (e.g., a video or image in a recommendation system), with each carrying a \textbf{\emph{feature vector}}. Edges between nodes capture their closeness relationship, which can be measured through a metric distance (e.g., Euclidean). There are a few main differences between the best-first traversal and classic BFS (breadth-first search) and DFS (depth-first search) algorithms. The first is an \textbf{\emph{ordering-based expansion}}. During graph traversal, the algorithm selects the closest unchecked node $v_i$, called an \textbf{\emph{active node}}, and computes the distance of all neighbors of $v_i$ to the query with their feature vectors (Line~\ref{algo_line:SGS_expand_1}-\ref{algo_line:SGS_expand_2}), and \emph{only} inserts promising neighbors into a \emph{priority queue} as new unchecked candidates for future expansion. In this way, the search can limit the number of distance computations needed to converge to near neighbors. Second, different from the BFS and DFS, which traverse all the connected nodes, the best-first search \textbf{\emph{converges}} when no new (unchecked) vertex can be found to update the priority queue, leading to a different number of convergence iterations (i.e., the number of while loop iterations in Algorithm~\ref{algo:seq_greedy_search}) for different datasets and queries.

\input{text/algo_seq_greedy_search}

\subsection{Complexities for Optimizations}

The graph traversal process in similarity graphs shares some common complexities with traditional graph processing for performance optimizations, but it also owns some distinctive features. However, no previous work has given a systematic examination of these complexities. Such knowledge is essential for optimizing similarity graph search, especially at a large scale.

\begin{figure}[t]
    \centering
    \includegraphics[width=0.45\textwidth]{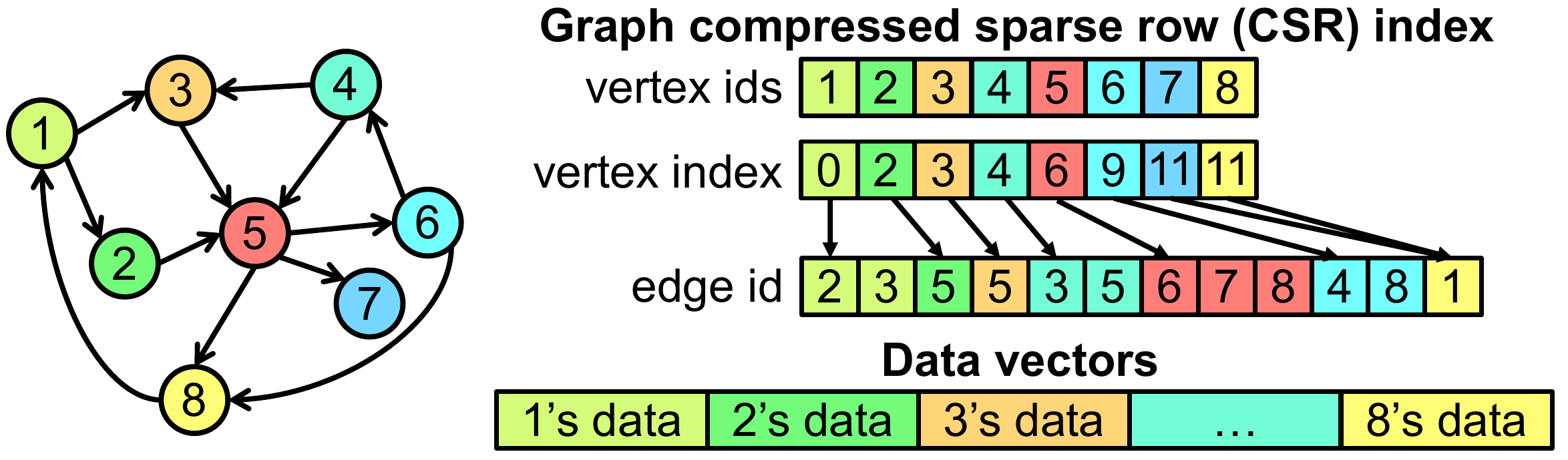}
    \caption{The storage structure of the graph-based index. 
        \textmd{The graph topology is stored in compressed sparse row (CSR) format, and the data vectors are stored in consecutive arrays.}}
    \label{fig:fig_csr_index}
\end{figure}

\noindent{\bf Challenge I: \SeqFullName (\SeqShortName) takes long iterations to converge, resulting in a prolonged critical path with heavy control dependency.}
As Algorithm~\ref{algo:seq_greedy_search} shows, 
this search consists of a sequence of search steps (Line~\ref{algo_line:SGS_while_1}-\ref{algo_line:SGS_while_2}) in which the candidates in the current step are determined by the last step.
Consider that ANNS usually queries for the top K nearest neighbors, requiring the first K elements in the priority queue to become stable. This state update usually converges slowly (e.g.,  $>$ 400 search steps or convergence steps to find the 100-nn with 0.999 recall for a million-scale dataset SIFT1M), resulting in a long critical path of execution. 

\noindent{\bf Challenge II: Limited edge-wise parallelism in traversal and memory bandwidth under-utilization.}
%Obvious parallelization is on the neighbor level, which is too fine-grained.
Beyond the aforementioned long convergence steps, it is possible to parallelize the neighbor expansion step (Line~\ref{algo_line:SGS_expand_1}-\ref{algo_line:SGS_expand_2} in Algorithm~\ref{algo:seq_greedy_search}) to reduce the execution time by dividing the neighbors into disjoint subsets and having multiple threads each compute the distance for a subset in parallel, which is called \emph{edge-wise parallelism}. 
However, this parallelism strategy often achieves sub-optimal performance, because many  similarity graphs have a small truncated out-degree on all nodes to avoid the \emph{out-degree explosion problem}~\cite{fu2019fast}. As a result, dividing the work across more worker threads would result in each thread processing only a very small number of vertices. Furthermore, edge-wise parallelism also adds synchronization overhead (e.g., at Line 14) to maintain an ordered expansion. Our preliminary experiment results in Table~\ref{tab:nsg_parallel_bdw} show that the edge-wise parallelism strategy (e.g., 
running with 64 threads on five datasets) leads to less than 5\% of the peak hardware memory bandwidth ($\sim$80 GB/s),
indicating a large performance potential remains yet to tap into.

\begin{table}[ht!]
\caption{Memory bandwidth (bdw.) measurement for edge-wise parallelism strategy.}
\label{tab:nsg_parallel_bdw}
\scalebox{0.84}{
\begin{tabular}{|c|c|c|c|c|c|}
    \hline
    Datasets         & SIFT1M & GIST1M & DEEP10M & SIFT100M & DEEP100M \\ 
    \hline \hline
    bdw. (GB/s) & 1.9    & 3.3    & 1.6     & 1.0        & 1.1      \\ \hline
\end{tabular}
}
\end{table}

\noindent{\textbf{Challenge III: Strict expansion order leads to high synchronization cost.}}
%All candidates are kept sorted in a priority queue, which makes simultaneous inserting very expensive even for the neighbor level parallelization.
Existing similarity graph search algorithms use a priority queue to maintain the strict priority order of all candidates according to their distances to the queue point. Although it is possible in principle to use a concurrent priority queue that uses locks or lock-free algorithms to synchronize the candidate insertions (Line 14),
we observe that the parallel scalability is severely limited by maintaining this strict order because each worker thread only performs distance computations for a few vertices.
% However, since each worker thread only execute distance computations for a few vertices. Therefore, it is imperative to make the concurrent insertions lightweight. However, we find that the parallel scalability is severely limited by maintaining this strict order. 

\noindent{\textbf{Challenge IV: Poor locality brought by irregular memory accesses.}}
Existing similarity graphs often store the graph index (e.g., in the compressed sparse row (CSR) format that contains a vertex array and an edge array) and feature vectors (e.g., in one embedding matrix) separately in memory as different objects, as shown in Figure~\ref{fig:fig_csr_index}. There are two points in this design that lead to inefficiencies. First, the accessed nodes often reside discontinuously in memory, which leads to unpredictable memory accesses. Second, it requires one-level of indirection to access feature vectors, leading to difficulties for memory locality optimizations.

%\begin{figure}[t]
%\centering
%\includegraphics[width=0.45\textwidth]{figures/fig_csr_index}
%\caption{The storage structure of the graph-based index. 
%    \textmd{The graph topology is stored in compressed sparse row (CSR) format, and the data vectors are stored in consecutive arrays.}}
%\label{fig:fig_csr_index}
%\end{figure}

%\minjia{Can we have some illustrative figure to show the irregular access and one-level of indirection issue?
%\zhenfix{Will it be similar to Figure~\ref{fig:fig_reorder}?}
%\minjia{Sort of. Figure~\ref{fig:fig_reorder} shows the optimized index. I am thinking that we can add a figure here so it is clear how the current index implementation looks like.}}

% \subsection{Deep Dive into the Compute Patterns of Best-First Search}

%% file: text/algo_seq_greedy_search.tex
\begin{algorithm}
\DontPrintSemicolon
\caption{\SeqFullName (\SeqShortName)}\label{algo:seq_greedy_search}
\KwIn{graph $G$, starting point $P$, query $Q$, queue capacity $L$}
\KwOut{$K$ nearest neighbors of $Q$}
priority queue $S$ $\gets \emptyset$\;
%set $S$'s capacity as $L$\;
index $i \gets 0$\;
compute $dist(P, Q)$\; \label{algo_line:SGS_starting_1}
add $P$ into $S$\; \label{algo_line:SGS_starting_2}
\While{has unchecked vertices in $S$} { \label{algo_line:SGS_while_1}
    $i \gets$ the index of the 1st unchecked vertex in $S$\; \label{algo_line:SGS_first_unchecked}
    mark $v_i$ as checked\;
    %    \verb|/*| Expand $v_i$ \verb|*/|\;
    \tcc{Expand $v_i$}
    \ForEach{neighbor $u$ of $v_i$ in $G$} { \label{algo_line:SGS_expand_1}
        \If{$u$ is not visited} {
            mark $u$ as visited\;
            compute $dist(u, Q)$\; \label{algo_line:SGS_adding_1}
            add $u$ into $S$\; \label{algo_line:SGS_adding_2}
        }
    } \label{algo_line:SGS_expand_2}
    \textbf{if} $S$.size() $> L$, \textbf{then} $S$.resize($L$)\;
} \label{algo_line:SGS_while_2}
\Return the first $K$ vertices in $S$\;
\end{algorithm}

%%%%%%%%%%%%%
%%% backup
%\begin{algorithm}
%\caption{\SeqFullName (\SeqShortName)}\label{algo:seq_greedy_search}
%\begin{algorithmic}[1]
%\REQUIRE graph $G$, starting point $P$, query $Q$, queue capacity $L$
%\ENSURE $K$ nearest neighbors of $Q$
%\STATE priority queue $S$ $\gets \emptyset$
%\STATE set $S$'s capacity as $L$
%\STATE index $i \gets 0$
%\STATE compute $dist(P, Q)$ \label{algo_line:SGS_starting_1}
%\STATE add $P$ into $S$ \label{algo_line:SGS_starting_2}
%%\WHILE{$i < L$} \label{algo_line:SGS_while_1}
%\WHILE{has unchecked vertices in $S$} \label{algo_line:SGS_while_1}
%    \STATE $i \gets$ the index of the 1st unchecked vertex in $S$
%    \STATE mark $v_i$ as checked
%%    \COMMENT{Expand $v_i$}
%    \STATE \verb|/*| Expand $v_i$ \verb|*/|
%    \FOR{every neighbor $u$ of $v_i$ in $G$} \label{algo_line:SGS_expand_1}
%        \IF{$u$ is not visited} 
%            \STATE mark $u$ as visited
%            \STATE compute $dist(u, Q)$ \label{algo_line:SGS_adding_1}
%            \STATE add $u$ into $S$ \label{algo_line:SGS_adding_2}
%        \ENDIF
%    \ENDFOR \label{algo_line:SGS_expand_2}
%\ENDWHILE \label{algo_line:SGS_while_2}
%\RETURN the first $K$ vertices in $S$
%\end{algorithmic}
%\end{algorithm}
%%% end backup
%%%%%%%%%%%%%

%% file: text/designv2.tex
\section{Design of ${\Hammer}$}
\label{sec:design}

Based on the observations from Section~\ref{subsec:speculation}, we introduce \Hammer, a
parallel search algorithm that exploits lightweight intra-query parallelism (i.e., path-wise parallelism and edge-wise parallelism) to accelerate the search efficiency of similarity graphs on multi-core CPU architectures. We first provide an overview of our architecture-aware design, and then we discuss technical details.

\begin{figure}[t]
    \centering
    \includegraphics[width=0.45\textwidth]{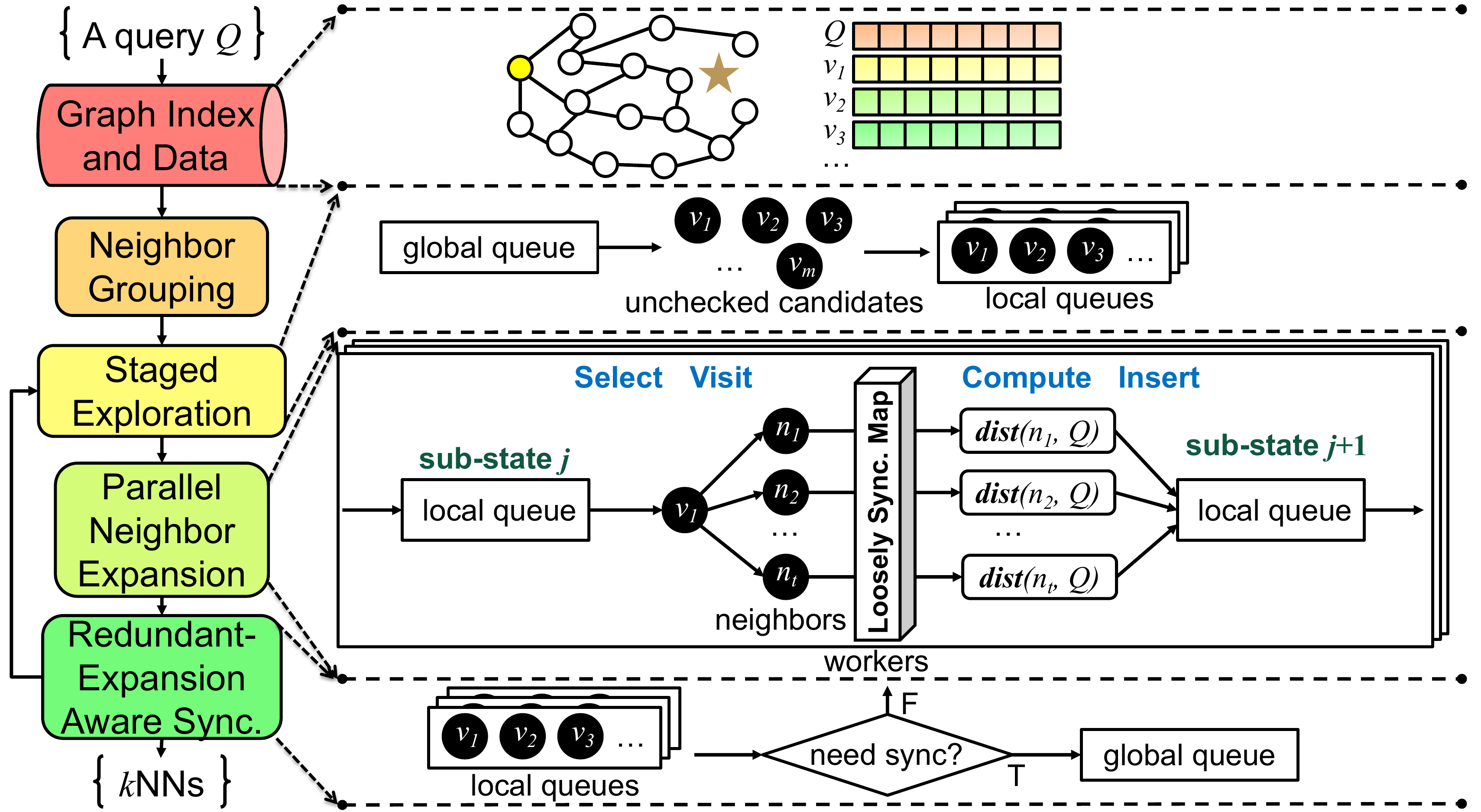}
    \caption{Overview of \Hammer.}
%            \emph{Dispatcher}, \emph{Executer}, and \emph{Sync Decider} are optimized in \ScaleM and \FullName.}}
% \minjia{TODO: This figure needs some update. On the left hand side, "local-subsearch"-> "Parallel Neighbor Expansion". "Order inversion tolerant"-> "Redundant-Expansion Aware Synchronization".}
    \label{fig:fig_system_overview}
\end{figure}

Figure~\ref{fig:fig_system_overview} depicts \Hammer's overall design that addresses the challenges mentioned in Section~\ref{sec:overview} to perform an efficient similarity graph search.
To reduce the long critical path dependency (Challenge I) and increase the amount of parallelism, \Hammer uses \emph{parallel neighbor expansion} to deliver coarse-grained parallelism.
\Hammer further introduces a staged search strategy to reduce redundant computations caused by over-expansion during a parallel search.
To limit global synchronization overhead (Challenge III), \Hammer adopts \emph{redundant-expansion aware synchronization} to adaptively adjust synchronization frequency.
As such, \Hammer reduces the number of global synchronizations while still achieving high search accuracy. 
Besides, \Hammer uses loosely synchronized visit maps for lightweight communication and also performs the neighbor grouping technique to improve memory locality (Challenge IV). 
% We next explain each of these techniques in detail.

\subsection{Parallel Neighbor Expansion} 
% \label{subsec:loosely_sync}
\label{subsec:speculation}

\begin{figure}[t]
    \centering
    \begin{subfigure}[t]{0.23\textwidth}
        \centering
        \includegraphics[width=\textwidth]{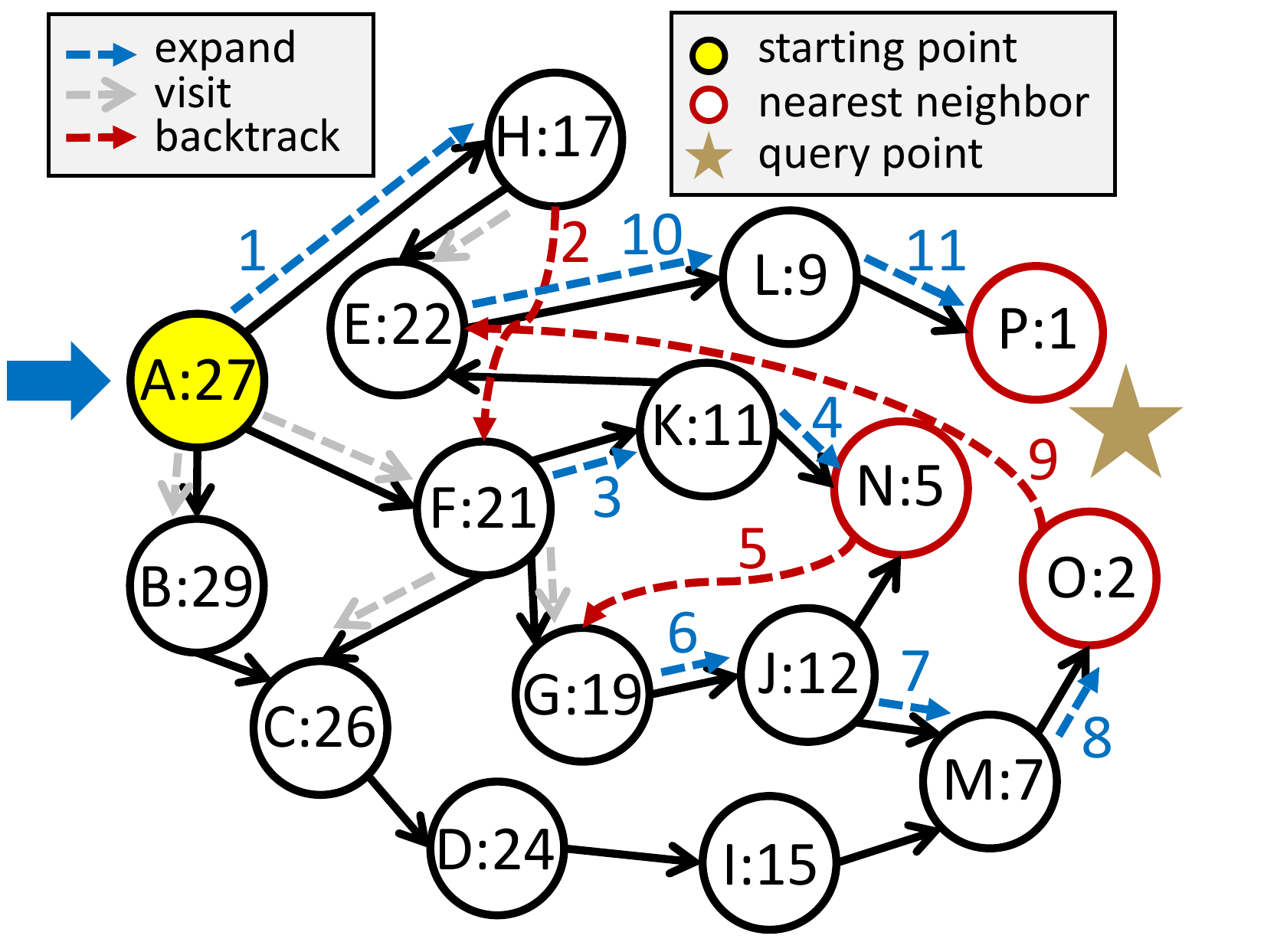}
        \caption{\SeqFullName w/ backtrack.}
        \label{subfig:search_path_A}
    \end{subfigure}
    \hfill
    \begin{subfigure}[t]{0.23\textwidth}
        \centering
        \includegraphics[width=\textwidth]{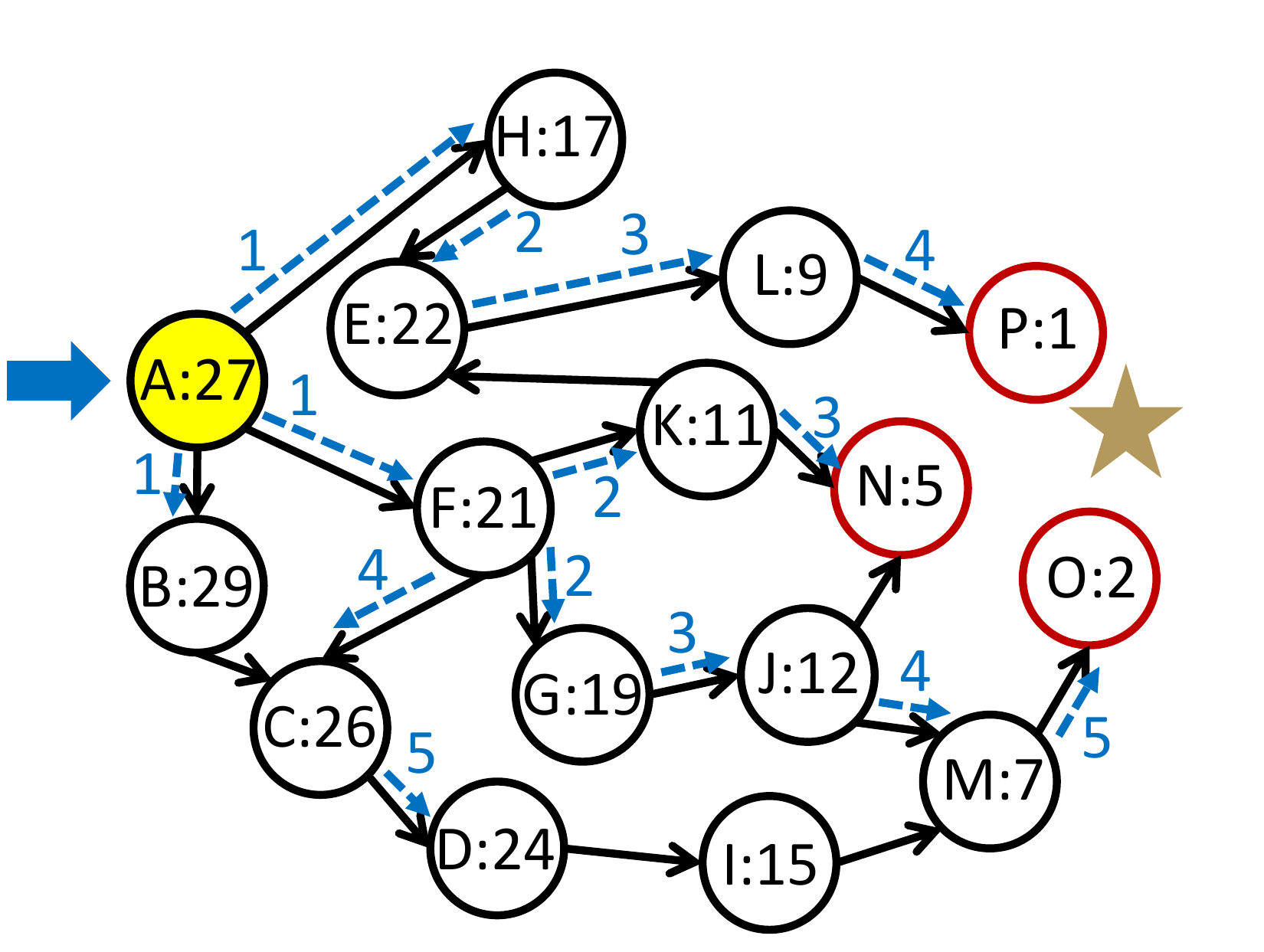}
        \caption{\TopM: expand top-3 candidates.}
        \label{subfig:search_path_B}
    \end{subfigure}
    \caption{Comparison of BFiS and \Hammer.
            \textmd{BFiS needs a long search path with backtrack to find nearest neighbors (11 steps). \Hammer reduces backtrack and completes with a shorter path (5 steps).}}
    \label{fig:example_search_path}
%    \vspace{-2em}
\end{figure}

Although it is challenging to parallelize the \SeqFullName (\SeqShortName) process due to its long critical path and limited edge-wise parallelism, the semantics of the algorithm does not seem to always require a strict order as long as the goal is to minimize the total search time of near neighbors. In this section, we exploit whether the search is robust to deviation from a strict order by allowing concurrent expansion of multiple active nodes. 
For practical similarity search, e.g., NSG and HNSW, there is no guarantee that a \emph{monotonic search path} always exists for any given query~\cite{fu2019fast}. As a result, the search can easily get trapped into the local optimum. To address this issue, \SeqShortName may \emph{backtrack} to visited nodes and find another out-going edge that has not been expanded to continue the search.
Figure~\ref{fig:example_search_path}(a) illustrates a search path with backtracking. The search starts from vertex $A$ and calculates the distance (indicated by the number following the letter on each vertex) between the three neighbors of $A$ ($B$, $F$, and $H$) and the query point. Because $H$'s distance is locally the smallest, \SeqShortName would select $H$ as the active node in the next step. However, given that further expanding $H$ no longer leads to a closer candidate, the search reaches a local minimum and performs a \emph{backtracking} to the next promising candidate $F$. The search process then may backtrack multiple times until it either finds the near neighbor (e.g., $O$) or exhausts the search budget.   

%%%% Introduce \TopM as shown in Algorithm~\ref{algo:par_top_m_search}.
Backtracking creates additional dependencies in \SeqShortName process and increases the convergence steps to find near neighbors. However, many of these backtracking dependencies can be \textbf{"fake" dependencies} if we perform a \emph{parallel neighbor expansion}, e.g., {\em by expanding multiple active nodes concurrently, it is possible to shorten the convergence steps by 
starting early at one of those backtracking points}. 
As an example, while it takes 11 steps to find the near neighbor in Figure~\ref{fig:example_search_path}(a), it only takes 5 steps in Figure~\ref{fig:example_search_path}(b) if we expand nodes F, G, J, M right after expanding their parent nodes.

Based on this insight, we introduce \Hammer. In this scheme, the priority order is relaxed such that in each step, {\bf top $M$ unchecked candidates} are selected as active nodes for expansion instead of just the best candidate. 

\begin{figure}[t]
    \begin{subfigure}[t]{0.23\textwidth}
        \centering
        \includegraphics[width=\textwidth]{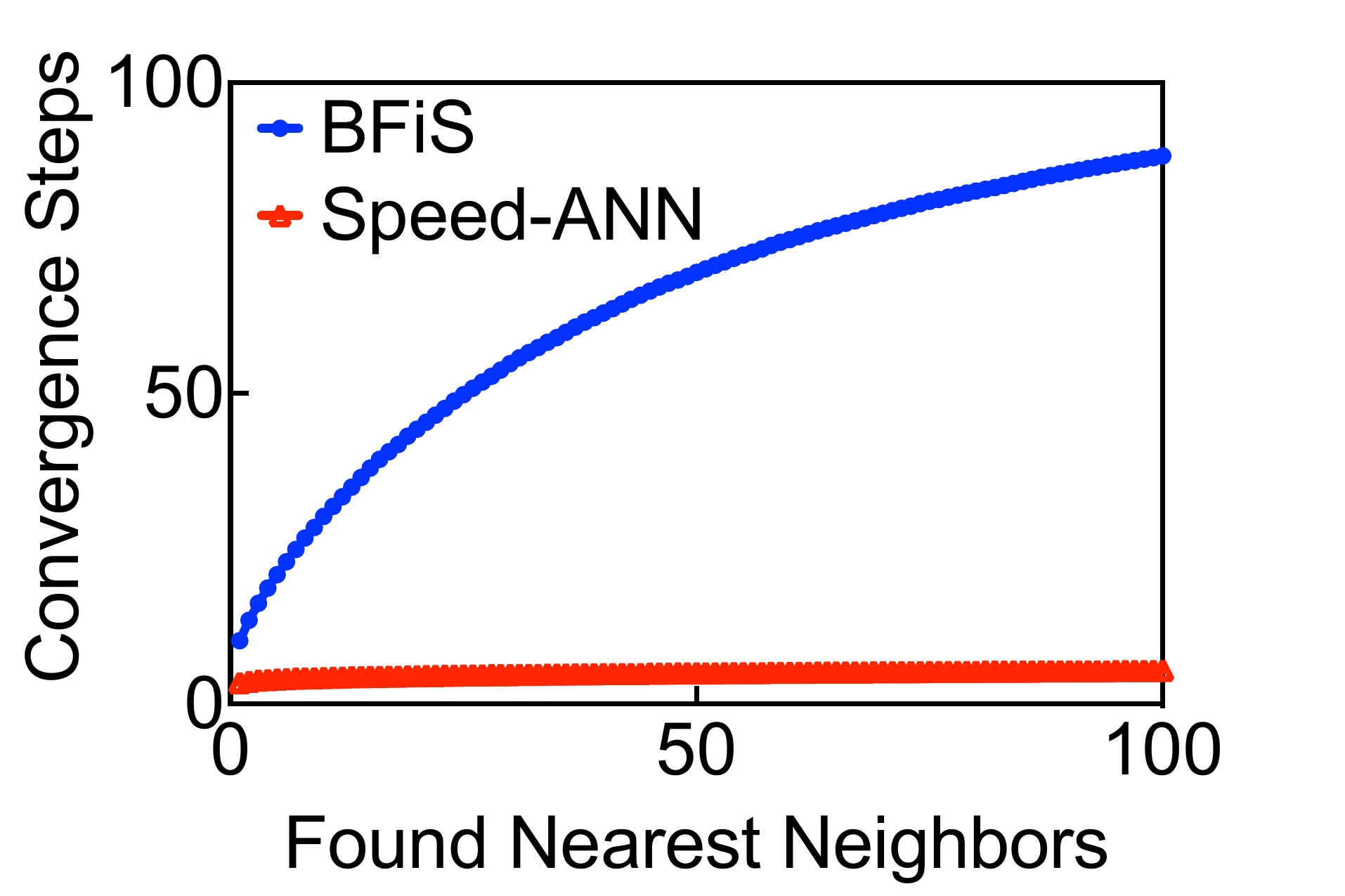}
        \caption{Convergence steps to find the $K$-th nearest neighbor in the queue. $K$ is specified by the x-axis.
            \textmd{Although \SeqShortName can find the first neighbor quickly, it still needs many steps to find all others.}}
        \label{subfig:insight_last_update_iter_vs_rank}
    \end{subfigure}
    \hfill
    \begin{subfigure}[t]{0.23\textwidth}
        \centering
        \includegraphics[width=\textwidth]{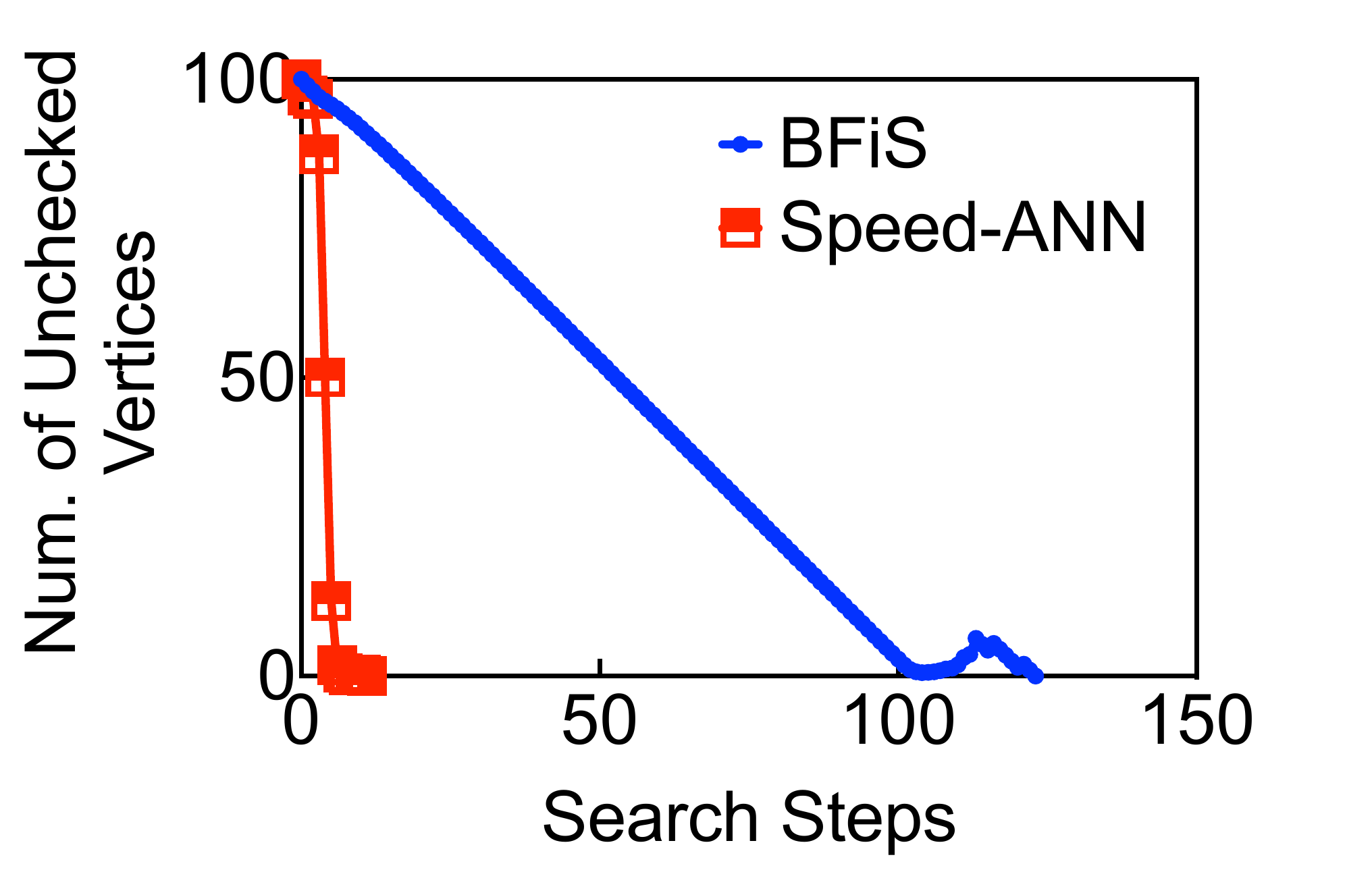}
        \caption{Numbers of unchecked candidates (vertices) in the queue after every search step. 
            \textmd{While \SeqShortName needs 100+ steps to converge, \TopMShortName only needs 10+ steps.
                Values are the average of 10K queries.}
        }
        \label{subfig:insight_unchecked_vs_iters}
    \end{subfigure}
    \caption{\TopMShortName results in much less search steps than \SeqShortName. 
        \textmd{
            Dataset is SIFT1M. They have the same $L=100$. \TopMShortName has $M=64$, where $M$ means the top $M$ unchecked candidates.}
    }
%    \minjia{Update the legend of these figures, e.g., MPS -> iQANS.}
    \label{fig:insight_convergence_steps_Top_M_vs_SGS}
    % \vspace{-2em}
\end{figure}

\begin{figure}[t]
    \begin{minipage}[t]{0.23\textwidth}
        \centering
        \includegraphics[height=0.98in]{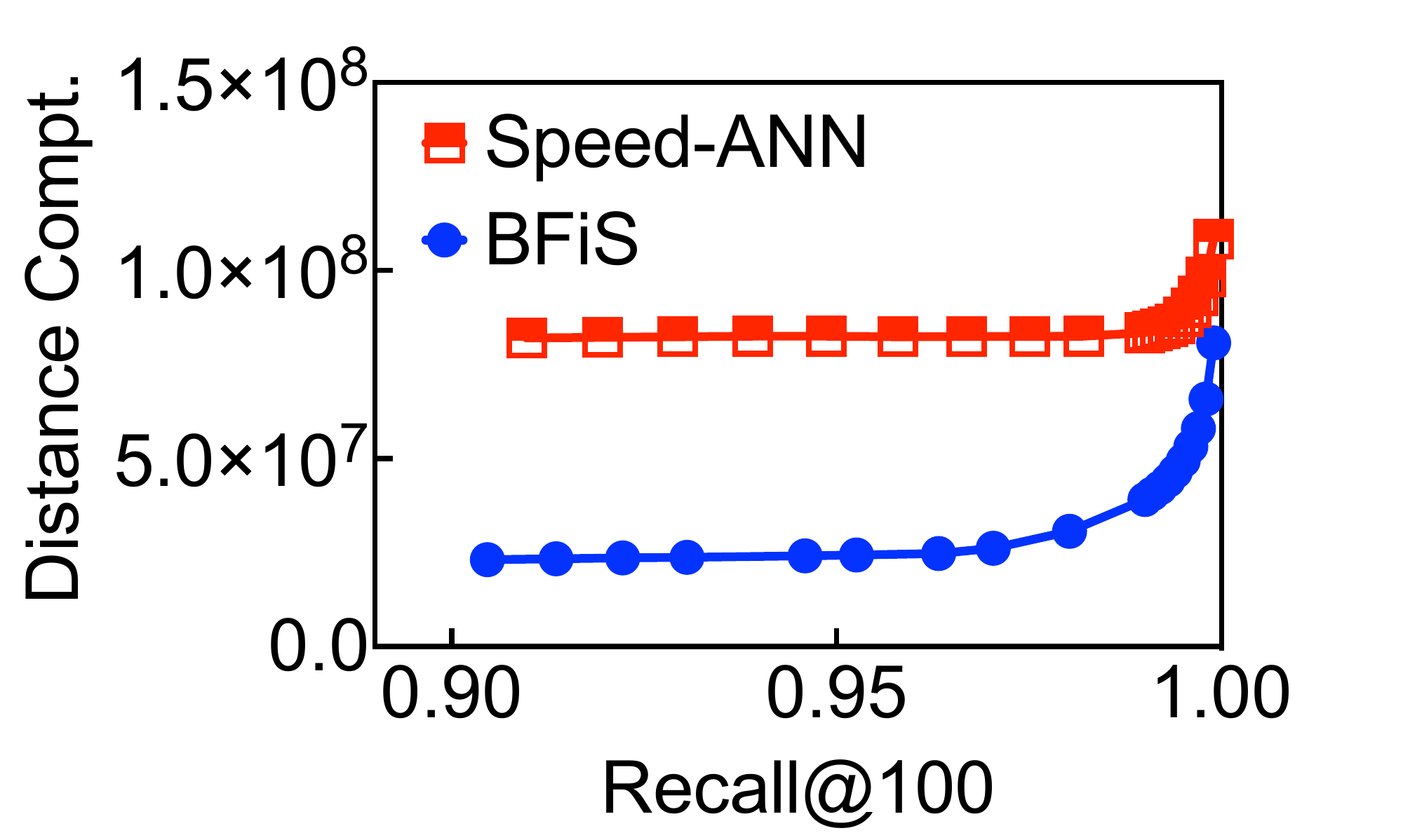}
        \caption{Distance computations of \SeqShortName and \TopMShortName, where $M = 64$.}
        \label{fig:insight_1T_compt_Top_M_vs_SGS}
        % \vspace{-4em}
    \end{minipage}
    \hfill
    \begin{minipage}[t]{0.23\textwidth}
        \centering
        \includegraphics[height=0.98in]{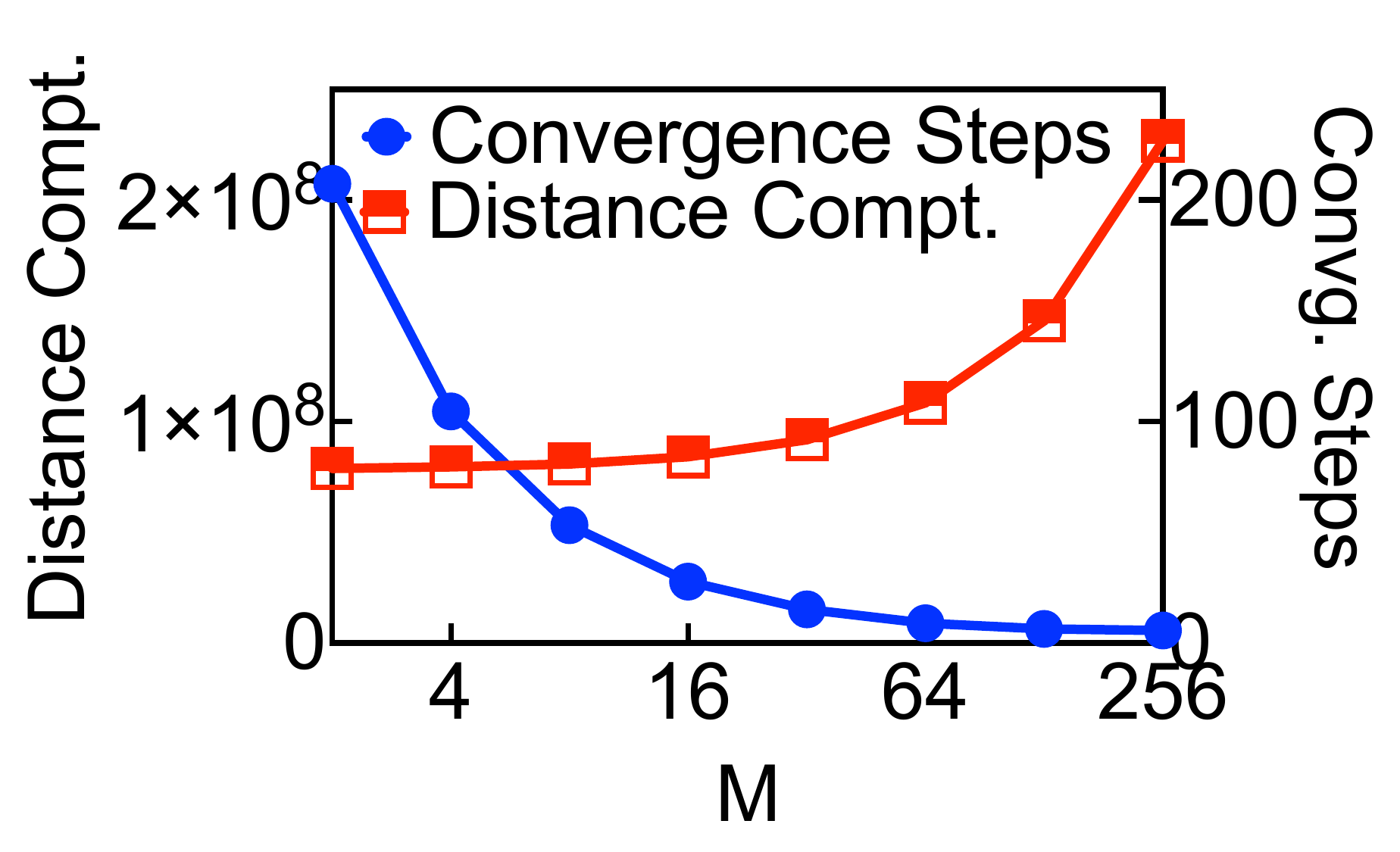}
        \caption{Distance computations and search steps of \TopMShortName when $M$ changes.
        }
        \label{fig:insight_Top_M_compt_steps_vs_M}
        % \vspace{-4em}
    \end{minipage}
%    \minjia{Where are these figures mentioned in the text?}
\end{figure}

\PunchStarter{\Hammer exposes hidden parallelism.}
The relaxation of the order enables two levels of parallelism: the \emph{path-wise parallelism} where multiple threads can concurrently expand the search frontier,
and the \emph{edge-wise parallelism} when expanding an individual active node. 
Moreover, instead of having a global queue to maintain strict expansion orders among all workers, each worker has a local priority queue,
which allows a thread to exploit a small number of \emph{order inversions} (i.e., allowing a worker thread to locally select and expand active nodes ahead of the global order), which can dramatically reduce communication, synchronization, and coordination between threads. 

\PunchStarter{\Hammer converges faster to near neighbors.}
One key benefit of \TopMShortName is that it significantly shortens the \emph{convergence steps} compared to \SeqShortName.
Figure~\ref{fig:insight_convergence_steps_Top_M_vs_SGS} shows the comparison results of convergence steps between \SeqShortName and \TopMShortName.
The results are measured on dataset SIFT1M using 10K queries with $0.90$ recall target, and $M$ is set to 64.
\TopM takes on average 3.4, 5.0, and 5.4 steps to find the 1st, 50th, and 100th nearest neighbor, respectively, whereas \SeqShortName takes 10.1, 69.4, and 88.1 steps, respectively.
From another aspect, \TopMShortName takes much fewer steps to finish examining all the unchecked vertices in $S$ than \SeqShortName, as shown in Figure~\ref{subfig:insight_unchecked_vs_iters}. Both results indicate that \TopMShortName has a much faster convergence speed than \SeqShortName.

\PunchStarter{Tree-based Expansion View.}
%To facilitate the further understanding of \TopMShortName, as well as its limitation and improvement, we introduce a Tree-based expansion view of \SeqFullName (\SeqShortName) and \TopM (\TopMShortName). 
Similar to the classical DFS/BFS, \SeqShortName naturally introduces an expansion tree: the root node $T_r$ of the tree is the starting vertex $P$ in graph $G$; the children of a tree node $T_i$ (corresponding to a graph vertex $v_i$) are the unvisited neighbors of $v_i$. The expansion of \SeqShortName bears many similarities to DFS, as each time, it will expand only one leaf node. However, different from DFS, which expands the one with the most depth, \SeqShortName expands the one which is closest to query $Q$. Thus, we have the same concepts of {\em backtracking} and {\em Steps} in \SeqShortName. 

The power of \TopM is that it expands the $M$ leaves simultaneously of the tree, which are $M$ nearest neighbors of query $Q$ among all the leaves of the {\em current} expansion tree. This effectively searches/extends $M$ paths in parallel instead of a single path (in \SeqShortName). Thus, \TopMShortName can potentially reduce the total number of {\em steps} of \SeqShortName by a factor of $M$ times, as for $k$ {\em Steps}, \TopMShortName can expand $k M$ tree nodes/leaves. Further, due to the hardware capability, at the same {\em time}, \TopMShortName can process $M$ leaves/paths expansion as only what is in \SeqShortName (one single leave or path expansion), leading to the low latency of query processing. 
We also note that the \SeqShortName becomes a special case of \TopMShortName where $M=1$, and both parallelization are under Bulk Synchronous Parallel (BSP) model~\cite{valiant1990bridging} though \SeqShortName has rather limited parallelism to explore.

% \begin{figure}[t]
%     \begin{minipage}[t]{0.23\textwidth}
%         \centering
%         \includegraphics[height=0.98in]{figures/insight_1T_compt_Top_M_vs_SGS}
%         \caption{Distance computations of \SeqShortName and \TopMShortName, where $M = 64$.}
%         \label{fig:insight_1T_compt_Top_M_vs_SGS}
%         % \vspace{-4em}
%     \end{minipage}
%     \hfill
%     \begin{minipage}[t]{0.23\textwidth}
%         \centering
%         \includegraphics[height=0.98in]{figures/insight_Top_M_compt_steps_vs_M}
%         \caption{Distance computations and search steps of \TopMShortName when $M$ changes.
%         }
%         \label{fig:insight_Top_M_compt_steps_vs_M}
%         % \vspace{-4em}
%     \end{minipage}
% %    \minjia{Where are these figures mentioned in the text?}
% \end{figure}

\subsection{Staged \Hammer to Avoid Over-Expansion}

% \PunchStarter{Challenges of \TopM.}
Despite the faster convergence speed, intra-query parallel search incurs additional challenges in increased distance computations. Figure~\ref{fig:insight_1T_compt_Top_M_vs_SGS} shows that to reach the same recall, \TopMShortName often leads to more distance computations than \SeqShortName. \TopMShortName has more computations because parallel neighbor expansion allows a query to take fewer steps to reach the near neighbors by avoiding fake dependencies from backtracking but it also introduces more computations to explore additional paths. Furthermore, we observe that although the convergence steps continue to decrease with larger $M$, the number of distance computations also increases dramatically, as shown in Figure~\ref{fig:insight_Top_M_compt_steps_vs_M}. 

\begin{figure}[t]
    %    \begin{minipage}[t]{0.22\textwidth}
        \begin{subfigure}[t]{0.23\textwidth}
            \centering
            \includegraphics[height=1.1in]{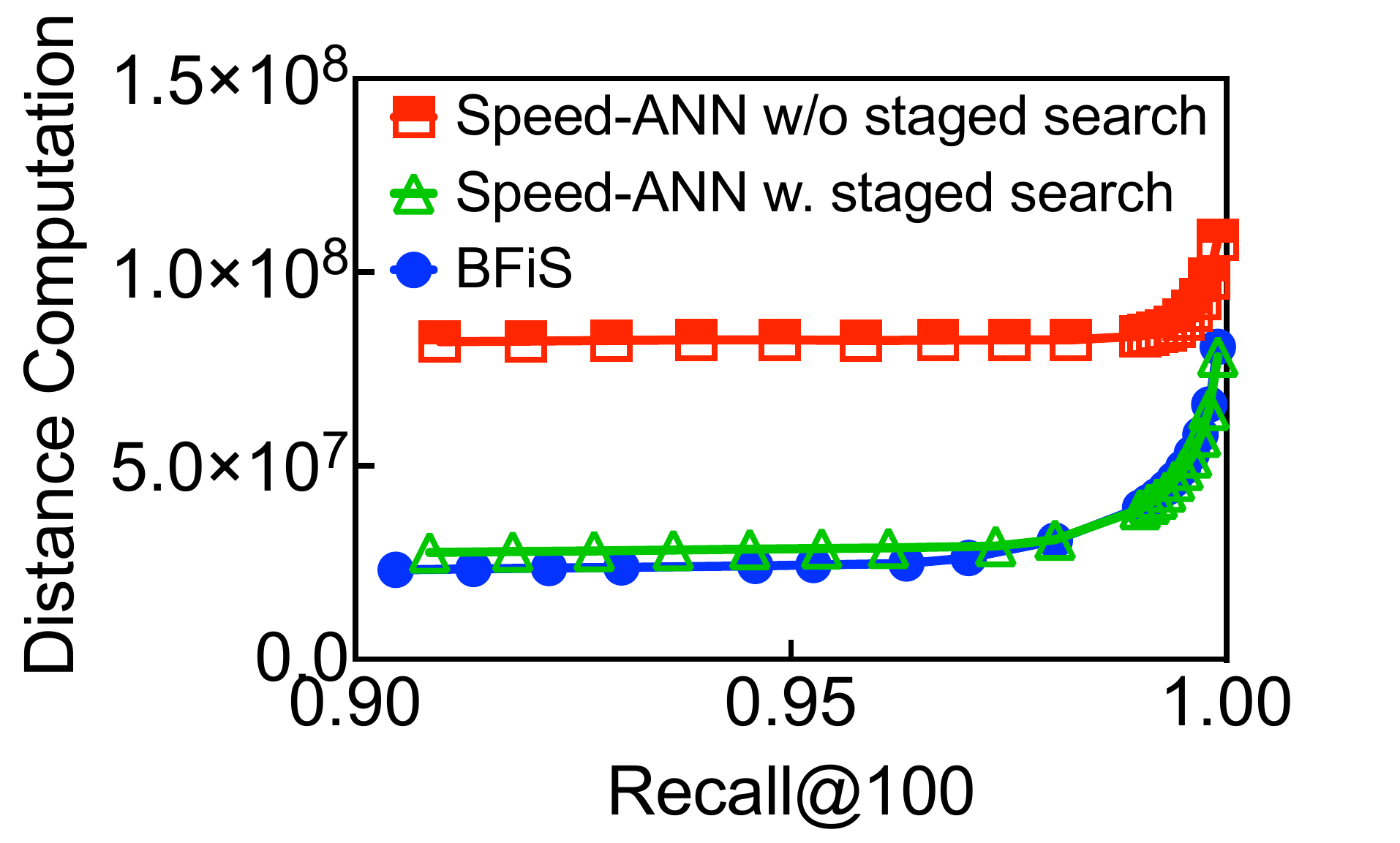}
            % \caption{Distance computation of \SeqShortName, \TopMShortName, and \ScaleMShortName.
            %     \textmd{\ScaleMShortName and \SeqShortName have similar distance computations.}}
            \caption{Distance computation of \SeqShortName, \Hammer w/o staged search, and \Hammer w. staged search.
                \textmd{Staged search avoids additional distance computation from over-expansion.}}
            \label{subfig:insight_1T_compt_Scale_M_vs_Top_M_vs_SGS}
            %    \end{minipage}
    \end{subfigure}
    \hfill
    %    \begin{minipage}[t]{0.22\textwidth}
        \begin{subfigure}[t]{0.23\textwidth}
            \centering
            \includegraphics[height=1.05in]{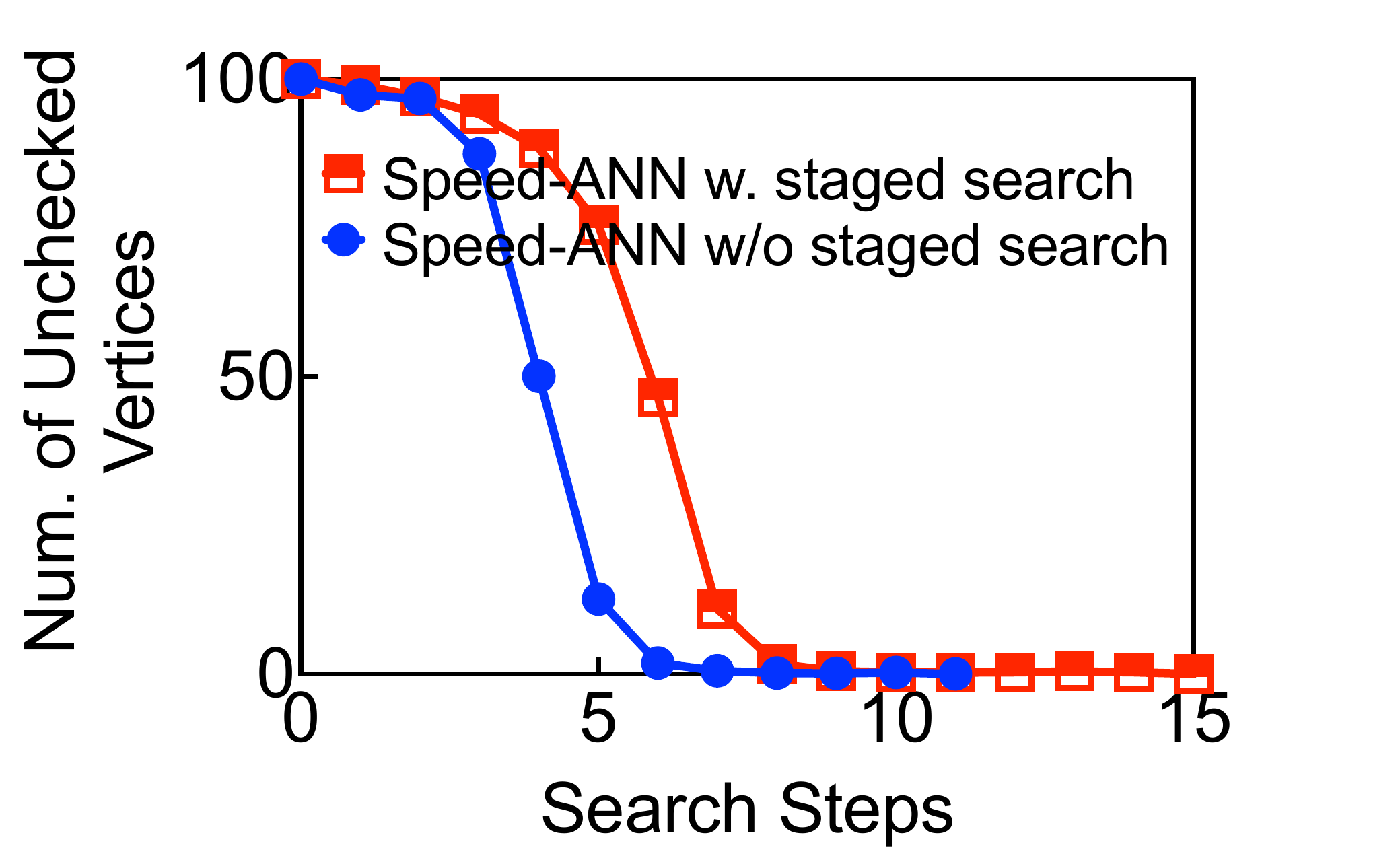}
            % \caption{\#unchecked candidates after each search step. 
            %     \textmd{\ScaleMShortName \& \TopMShortName have similar numbers  of steps.}}
            \caption{Number of unchecked candidates after each search step. 
                \textmd{\Hammer carries fast convergence properties.}}
            \label{subfig:insight_unchecked_vs_iter_Top_M_Scale_M}
            %    \end{minipage}
    \end{subfigure}
    \caption{Comparison between \Hammer without staged search and with staged search: distance computation \& search steps.
        \textmd{$M = 64$.}}
    \label{fig:insight_Top_M_vs_Scale_M}
%    \minjia{TODO: Zhen, can you update the legend in these two figures. MPS->"iQANS w/o staged search". P-MPS -> iQANS? In general, please go through the whole paper to make sure the names are consistent.}
    \vspace{1em}
\end{figure}

When the number of parallel workers is large, the search speed of \Hammer might be sluggish because the over-expansion of neighbors can result in many redundant computations during the entire search process. To avoid unnecessary distance computations caused by over-expansion, we take a \emph{staged} search process by gradually increasing the expansion width (i.e., M) and the number of worker threads every $t$ steps during the search procedure. The intuition is that the search is less likely to get stuck at a local minimum at the beginning of the search, so the best-first search with a single thread can already help the query to get close to near neighbors. As the search moves forward, it becomes more likely that a query will get stuck at a local minimum and requires backtracking to escape from the local minimum. Therefore, a parallel neighbor expansion search with a larger expansion width in later phases can better help reduce the convergence steps. We find that a simple staging function works well in practice: when the search begins, we first set a starting value and a maximum value for $M$. The starting value is usually one, and the maximum value can be as large as the number of available hardware threads. Subsequently, for every $t$ steps (e.g., $t=1$) we double the value of $M$ until $M$ reaches its maximum. 
Figure~\ref{subfig:insight_1T_compt_Scale_M_vs_Top_M_vs_SGS} shows that by taking staged search, \Hammer reduces the amount of redundant significantly in comparison to \Hammer without staged search and leads to distance computations close to \SeqShortName. On the other hand, \Hammer is able to converge as almost fast as \Hammer without staged search, as shown in 
Figure~\ref{subfig:insight_unchecked_vs_iter_Top_M_Scale_M}. These results indicate that our staged search method still achieves fast convergence speeds without incurring too many distance computations caused by over-expansion through the parallel search on a large number of workers. 

% In principle, if the underlying hardware has infinite compute and memory bandwidth, then these additional computations do not affect the search latency because they can be executed in parallel. In practice, the communication between cores is expensive and the memory bandwidth is not unlimited. 

% Moreover, the memory systems of multicores are hierarchical, where poor cache locality is still an concern due to the natural irregularity of the underlying graph.  
% In the next section, we propose an architecture-aware design that exploits intra-query parallelism together with multicores and memory hierarchy to maximize the scalability of search efficiency. 

\subsection{Redundant-Expansion Aware Synchronization}

\begin{figure}[t]
    \begin{minipage}[t]{0.23\textwidth}
        \centering
        \includegraphics[height=0.98in]{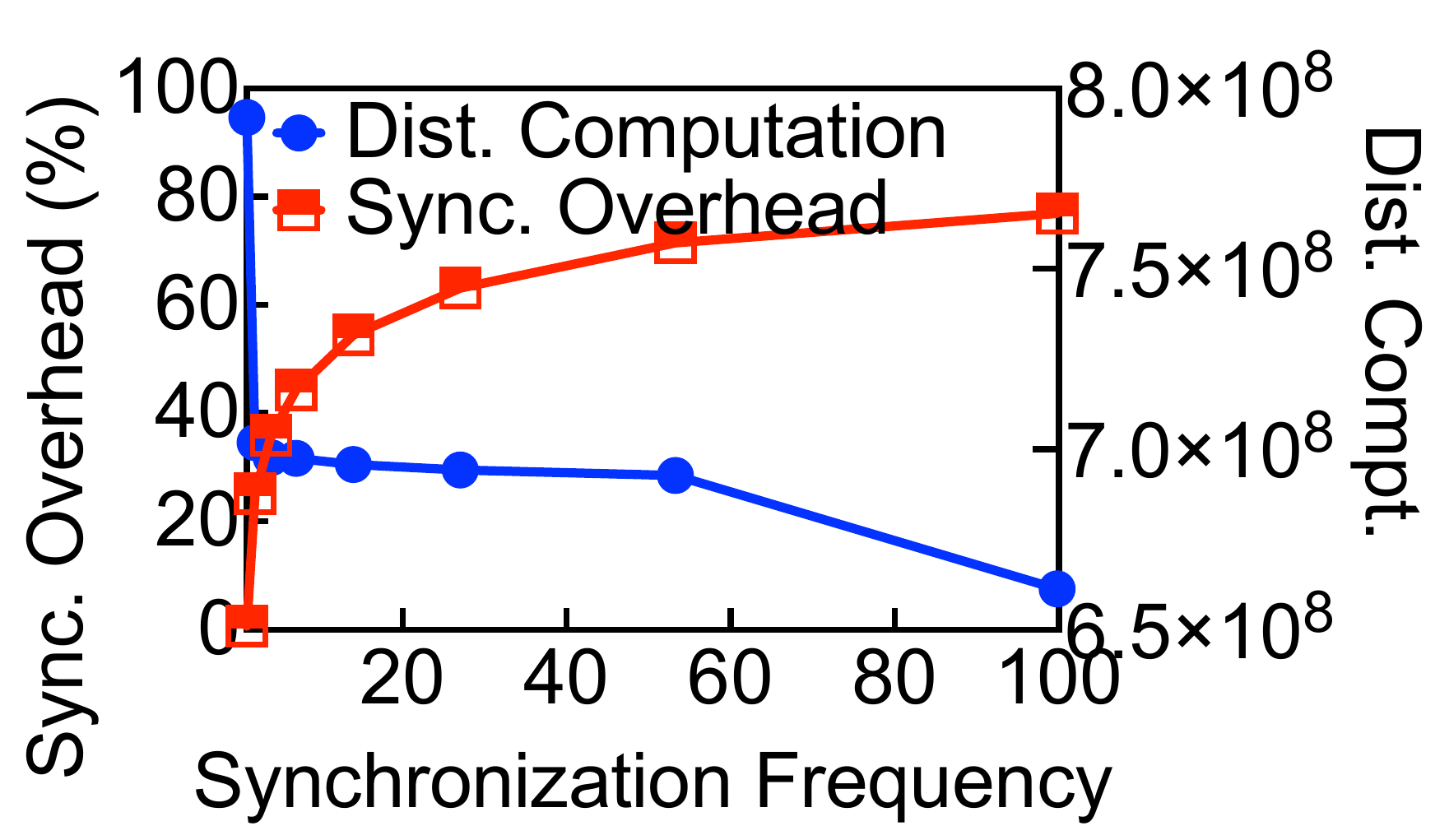}
        \caption{\Hammer's sync. overhead and distance computation vs. sync. frequency.}
        \label{fig:insight_PSS_sync_frequency_vs_overhead}
    \end{minipage}
    \hfill
    \begin{minipage}[t]{0.23\textwidth}
        %        \centering
        %        \includegraphics[height=0.98in]{figures/insight_PSS_sync_frequency_vs_ratio_1st_half}
        %        \caption{Ratio of the average update position in the 1st half of private queues.
            %            %\textmd{When $M$ is increasing, \TopMShortName's search steps are declining while its distance computation is increasing.}
            %        }
        %        \label{fig:insight_PSS_sync_frequency_vs_ratio_1st_half}
        \centering
        \includegraphics[height=0.98in]{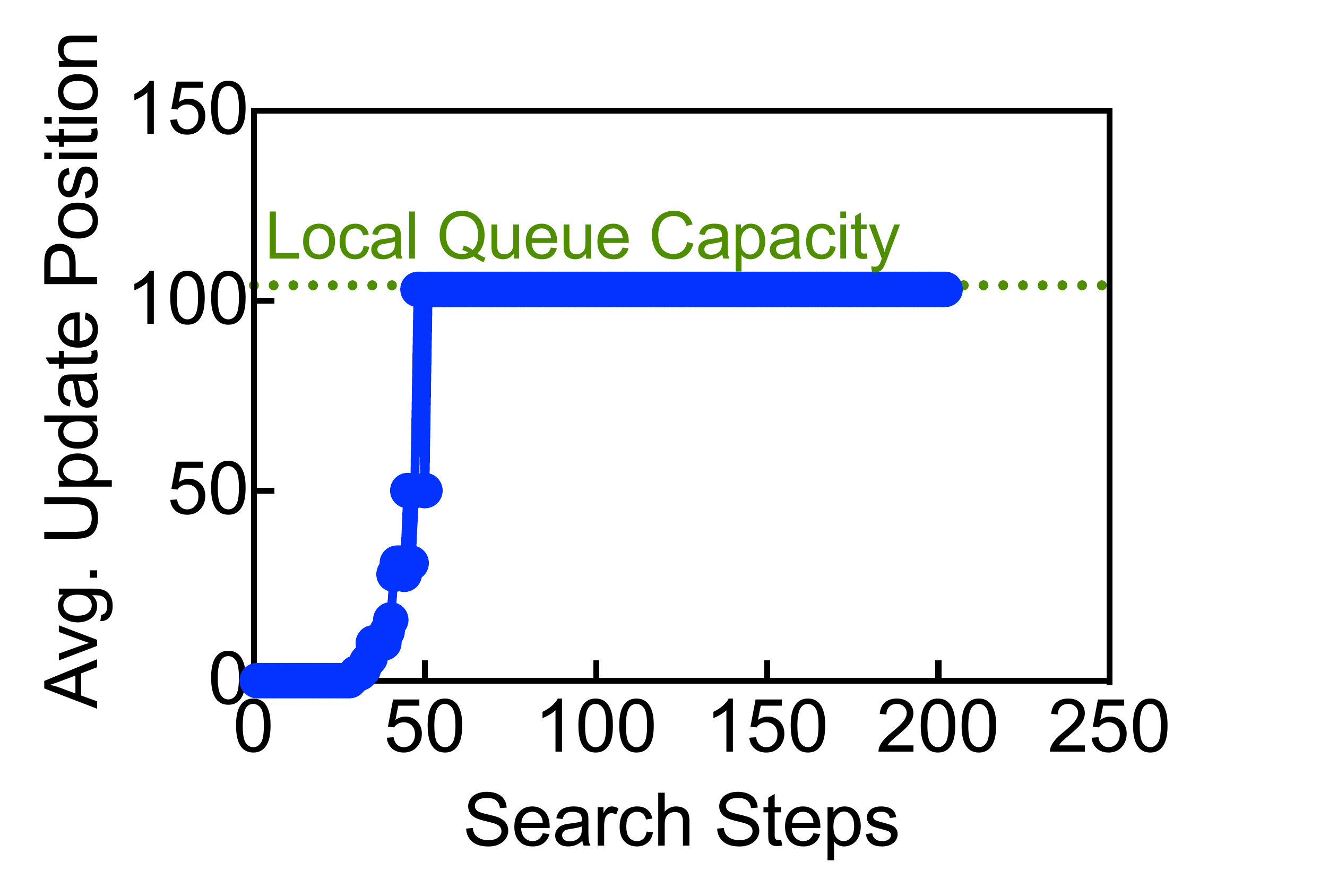}
        \caption{A query's average update positions during searching.
            %\textmd{When $M$ is increasing, \TopMShortName's search steps are declining while its distance computation is increasing.}
        }
        \label{fig:insight_PSS_update_position_example}
    \end{minipage}
    %    \caption{\TopM has much less search steps than \SeqFullName (\SeqShortName) does. The dataset is SIFT1M. They have the same $L=100$. \TopM has $M=64$.}
    %    \label{fig:insight_convergence_steps_Top_M_vs_SGS}
    \vspace{1em}
\end{figure}

As mentioned in Section~\ref{sec:overview}, yet another big performance bottleneck in intra-query parallelism resides in the synchronization overhead. Figure~\ref{fig:insight_PSS_sync_frequency_vs_overhead} shows how the global synchronization frequency influences the synchronization overhead (calculated by synchronization time divided by overall execution time) and the overall distance computations. All results in this figure return the same recall value. 
It shows that the synchronization overhead increases significantly when the synchronization frequency grows.
We also find that order inversion (without enough synchronization) slows down the search convergence and results in growing distance computations (as shown in Figure~\ref{fig:insight_PSS_sync_frequency_vs_overhead}). This is because, without enough synchronization, worker threads keep searching their own (unpromising) areas without benefiting from other threads' latest search results that may lead to faster convergence. This study demonstrates that a proper synchronization frequency is desired to achieve high system performance.

\noindent\textbf{Measuring redundant expansion via update positions.}
To unleash the full power of multi-core systems, \Hammer performs a unique form of lazy synchronization so that worker threads do not need to synchronize at every search step in most cases. 
Especially, our synchronization scheme is \emph{redundant-expansion aware}, which means instead of having a strict order through the entire convergence steps, we allow some relaxation of the order as long as each worker thread is still performing some effective search and the global order becomes consistent again after a large amount of redundant expansion has been detected.
In this paper, we propose a new way to measure the effectiveness of intra-query parallel search based on the \emph{update positions} of workers. When a worker expands an unchecked candidate, its neighbors are then inserted into the worker's local queue, and the update position is defined as the \emph{lowest (best)} position of all newly inserted candidates. Thus, \emph{the average update position} is the mean of all update positions of workers. 
Figure~\ref{fig:insight_PSS_update_position_example} demonstrates how an example query's average update position changes during the search steps without global synchronization. 
It shows that the average update position increases gradually to the local queue capacity and resides there to the end. 
When the average update position is close to the queue capacity, it indicates that most workers are searching among unpromising areas and cannot find good enough candidates to update their local results.
Therefore, the average update position can be used as a metric to determine if all workers need to synchronize their local results to adjust the search order.  We would like to note that there could be more than one metric to decide when to perform the lazy synchronization. We leave it as an open research question and more advanced methods might lead to better performance improvements. 

Algorithm~\ref{algo:check_merge_metric} describes how to use the average update position as the metric to decide when to perform a lazy synchronization. Given the queue capacity $L$ and a position ratio $R$, the threshold of the average update position to do synchronization is set as $L\cdot R$. 
If the \emph{checker} finds the average update position is greater than or equal to the threshold (Line~\ref{algo_line:check_metric}), it returns \texttt{true} indicating a global synchronization in Algorithm~\ref{algo:par_stale_search}.
Empirically, the ratio $R$ is close to $1.0$, such as $0.9$ or $0.8$. The input vector of all update positions is updated by workers regularly without locks. The return flag is only written by the \emph{checker} who is assigned among workers in a round-robin manner.

\input{text/algo_check_merge_metric}

Table~\ref{tab:comp_no_sync_bulk_step} shows preliminary results about the performance comparison between adaptive synchronization and no-synchronization. No-synchronization means each thread performs its local search and only combines the results in the end. The results show that adaptive synchronization is able to improve search efficiency with fewer distance computations. Overall, the reduced synchronization and distance computation from our redundant-expansion-aware synchronization is especially helpful for parallel neighbor expansion on a large number of workers, because global synchronization across multiple threads is still expensive and not very scalable as the number of cores increases. 

\begin{table}[!ht]
    \caption{Comparison between no-sync. and adaptive sync.
        \textmd{8 threads on SIFT1M for Recall@100 0.9. 
            Adaptive sync. check workers' dynamic status and merge queues adaptively.
            \texttt{Lt.} denotes latency. \texttt{Compt.} denotes distance computation.}}
    
    \label{tab:comp_no_sync_bulk_step}
    \begin{tabular}{|c|c|c|c|c|}
        \hline
        Dataset &               \multicolumn{2}{c|}{no-sync.}               &          \multicolumn{2}{c|}{adaptive sync.}           \\ \cline{2-5}
        &         Lt. (ms.)         &            Compt.            &         Lt. (ms.)         &           Compt.            \\ \hline\hline
        SIFT1M  & \multicolumn{1}{c|}{1.16} & \multicolumn{1}{c|}{125.3 M} & \multicolumn{1}{c|}{0.70} & \multicolumn{1}{c|}{33.1 M} \\ \hline
    \end{tabular}
\end{table}

\noindent
\textbf{Putting It Together.} 
Algorithm~\ref{algo:par_stale_search} describes the overall algorithm of \Hammer. At the beginning of each global step, the global queue evenly divides its unchecked candidates among all local threads.
After that, each worker performs a local best-first search based on its own local queue of sub-states (Line~\ref{algo_line:stale_substate_start} to Line~\ref{algo_line:stale_substate_end}).
Different from the global state that involves updating the global queue, a worker's local \emph{sub-state} is the state of its private queue.
In a local search step, a worker expands its own best unchecked candidate and updates its private queue accordingly.
Before the global queue's state is updated, a worker can have multiple sub-states of its own private queue.
%A worker continues expansion until it already has $I_{sync}$ sub-states 
A worker continues expansion until \texttt{CheckMetrics()} raises a flag for merging
or it has no unchecked candidates left locally. In a round-robin way, a worker is assigned as the \emph{checker}. 
Her duty is to check (as what \texttt{CheckMetrics()} does) if all workers need to synchronize their sub-states by merging all private queues into the global queue.
If so (Line~\ref{algo_line:checker_true}), all workers will stop their local search and merge their queues. 
% The definition of \texttt{CheckMetrics()} is flexible. 
% It could be a static method that no worker should do local search beyond a given step limit, or it could be an adaptive method that uses some metrics to determine workers' dynamic status.
% The details will be provided later in this section.
% \minjia{Move some part into impl.}

% As shown in Algorithm~\ref{algo:par_stale_search}, our synchronization frequency is controlled by \texttt{CheckMetrics()}, i.e., it determines when all workers merge their private queues into the global one. 
% A possible approach of \texttt{CheckMetrics()} is using exhaustive tuning method to find the proper search steps for global synchronization in terms of the shortest latency and recall guarantee. 
% However, this exhaustive method has a couple of disadvantages. 
% First, the search procedure for the ideal synchronization point is time-consuming. Given a query, the time complexity of finding the optimal setting of synchronization points is factorial about the total length of the search path.
% For a given dataset, this can only be reprocessed offline.
% Second, the tuning result is input-sensitive. Even a nuance in datasets or queries may result in total difference synchronization points which need be tuned exhaustively start over. Therefore, although the exhaustive method may provide ideal results, it is impractical.

\input{text/algo_par_stale_search}
\subsection{Additional Optimizations}

\noindent\textbf{Loosely Synchronized Visiting Map.}
There is one potential bottleneck to multi-threaded parallel scaling in Algorithm~\ref{algo:par_stale_search} on our target architectural platforms (multi-core systems). 
Consider visiting a neighbor of a candidate. This is typically after a check and then an update to a visiting map to ensure that a vertex is calculated once (Line~\ref{algo_line:stale_check_map}-\ref{algo_line:stale_update_map}).
During parallel neighbor expansion, the visiting map is shared by all workers to indicate if a vertex has been visited.
Since multiple threads may access the shared visiting map concurrently, locking or lock-free algorithms are required if we still want to ensure a vertex is visited only once. However, this approach involves a significant scalability bottleneck, because it leads to lock contention and sequentialization of updating the visiting map.

We observe that the ANN search algorithm is still correct even if a vertex is calculated multiple times because the local candidates are guaranteed to be merged back to the global priority queue and the visiting map is also guaranteed to have \emph{eventual consistency} the next time of global synchronization. Furthermore, by inserting memory fences, cache coherence further ensures that the updated visiting map is visible to other cores. Due to the potential out-of-order execution in processors, modern multi-core processors provide \emph{fence} instructions as a mechanism to override their default memory access orders. In particular, we issue a fence after a thread updates the visiting map to guarantee a processor has completed the distance computation of the corresponding vertex and has updated the visiting map (otherwise, there is no guarantee the updated visiting map is visible to other cores before next step of global synchronization).

By doing the loosely synchronized local search, we observe that the search algorithm only performs a very small percentage of additional distance computations (less than 5\%) for {SIFT1M} (and similar for other datasets) with 8-way parallelism. This reduces  the overhead from synchronization by 10\% and allows us to avert the issue of non-scaling locking across the multi-threading search. This optimization was also considered by Leiserson and Schardl~\cite{benign-race} (termed as "benign races") for their parallel breadth-first search algorithm. 
Furthermore, we use a bitvector to implement the visiting map instead of a byte-array.
% where byte is the minimum data type with Compare-and-Swap (CAS) support. 
This optimization allows the cache to hold the largest possible portion of the visiting map and therefore improves the data locality for memory accesses. 

\begin{figure}
    \centering
    \includegraphics[width=0.45\textwidth]{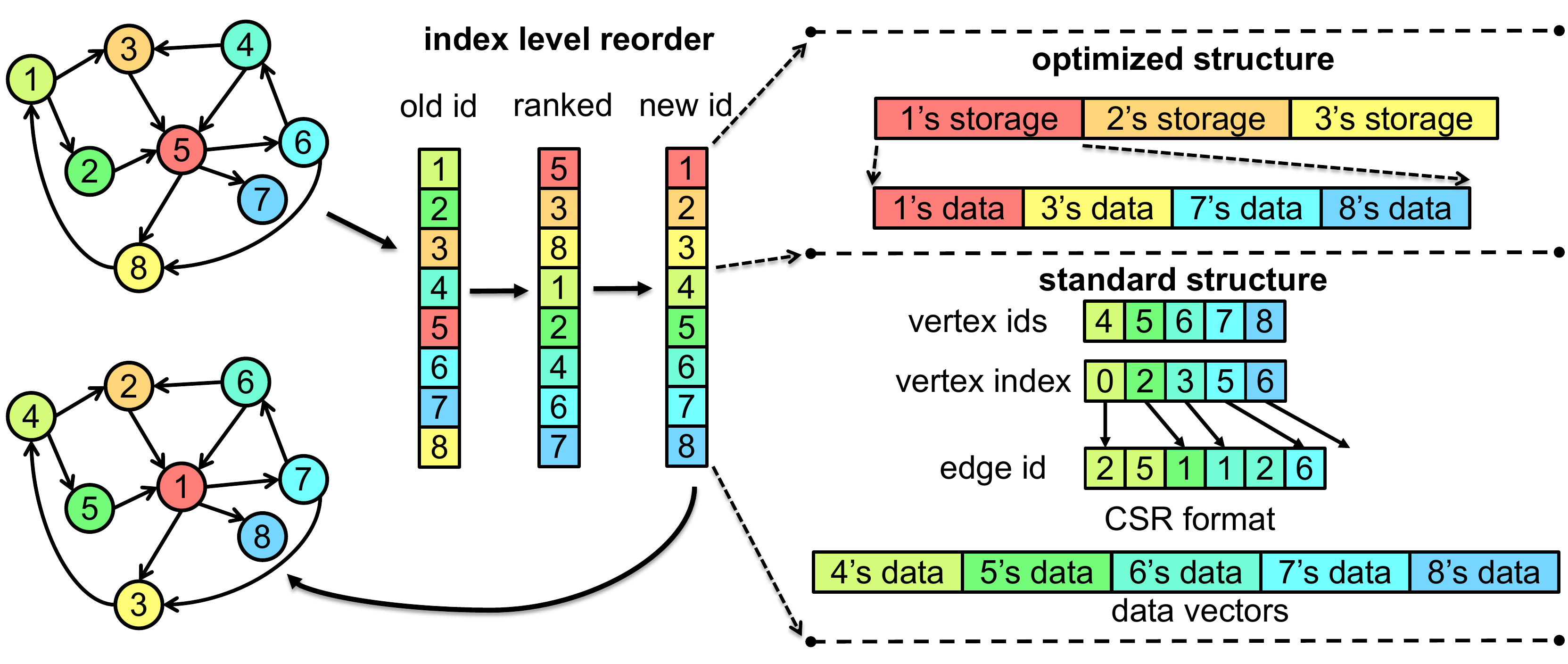}
    \caption{Example of neighbor grouping and hierarchical data storage. 
            \textmd{
                    Vertices are ranked according to their in-degree. 
                 Vertices are first reordered into new ids according to their ranks.
                    High ranked vertices are stored in an optimized index where every vertex's neighbors' data are stored in consecutive locations right after its own data to make expanding cache-friendly. 
                    Other low ranked vertices are stored in a standard index where the graph index and data vectors are stored separately.}}
    \label{fig:fig_reorder}
    % \vspace{1em}
\end{figure}

\noindent{\textbf{Cache Friendly Neighbor Grouping.}}
When a feature vector is loaded into memory for distance computation, modern CPU architectures actually automatically load vectors from nearby memory locations as well. 
Our neighbor grouping technique taps into this feature to mitigate the two levels of irregularity mentioned in Section~\ref{sec:overview}.

First, \Hammer \emph{flattens} the graph indices by placing the embeddings of neighbor vertices in consecutive memory, which would avoid one-level of implicit memory addressing and enables a thread to pre-fetch neighbor feature vectors once an active node is selected.  Second, \Hammer also regroups nodes, such that vertices that are likely to be visited during the graph traversal are already pre-load into the CPU memory and cache. Together, these two optimizations increase the cache hit rate and help speed up the search process.

One caveat of this approach is that it introduces additional memory consumption, because two neighbor lists may share the same vertex as a common neighbor. It is therefore may require more memory consumption than the original approach.
To avoid increasing the memory consumption, \Hammer takes a hierarchical approach by regrouping only a subset of vertices. In particular, \Hammer divides a graph to a two-level index as shown in Figure~\ref{fig:fig_reorder}, where only the top-level vertices have their neighbors flattened and stored in consecutive memory, and the bottom-level index stores other vertices using the standard structure. 
In this work, we explore two strategies to graph division: 
     \textbf{Degree-centric}, which puts high in-degree nodes to the top-level of the indices. The intuition is that high in-degree nodes are more frequently accessed, and therefore improving their locality would benefit the most for the overall search efficiency.
     \textbf{Frequency-centric}, which exploits query distribution to figure out which nodes are more frequently accessed and puts those frequently accessed nodes into an optimized index.
Section~\ref{sec:eval} evaluates both strategies and shows that \Hammer's neighbor grouping strategy brings 10\% performance improvements with selecting only 0.1\% vertices as the top level for a dataset with 100M vertices.

%% file: text/algo_check_merge_metric.tex
\begin{algorithm}
\DontPrintSemicolon
\caption{CheckMetrics() (Update Position Version)}\label{algo:check_merge_metric}
\KwIn{vector of update positions $U$, queue capacity $L$, position ratio $R$, number of workers $T$}
\KwOut{true or false}
$\bar{u} \gets$ average positions of elements in $U$\;
\If{$\bar{u} \geq L \cdot R$} { \label{algo_line:check_metric}
    \Return true\;
}
\Else {
    \Return false\;
}
\end{algorithm}

%%%%%%%%%%%%%
%%% backup
%\begin{algorithm}
%\caption{CheckMetrics() (Update Position Version)}\label{algo:check_merge_metric}
%\begin{algorithmic}[1]
%    \REQUIRE  vector of update positions $U$, queue capacity $L$, position ratio $R$, number of workers $T$
%    \ENSURE true or false
%    \STATE $\bar{u} \gets$ average positions of elements in $U$
%    \IF{$\bar{u} \geq L \cdot R$}
%        \RETURN true
%    \ELSE 
%        \RETURN false
%    \ENDIF
%\end{algorithmic}
%\end{algorithm}
%%% backup
%%%%%%%%%%%%%

%% file: text/algo_par_stale_search.tex
\begin{algorithm}
\DontPrintSemicolon
%\caption{\FullName (\ShortName)}\label{algo:par_stale_search}
\caption{\Hammer Intra-Query Parallel ANN Search}\label{algo:par_stale_search}
\KwIn{graph $G$, starting point $P$, query $Q$, queue capacity $L$, number of workers $T$}
\KwOut{$K$ nearest neighbors of $Q$}
expansion width $M \gets 1$\;
%global priority queue $S$ $\gets$ an empty queue with capacity $L$\;
%local priority queues $LS$ $\gets$ $T$ empty queues with capacity $L$\; \label{algo_line:stale_local_queues}
global priority queue $S$ $\gets$ an empty queue\;
local priority queues $LS$ $\gets$ $T$ empty queues\; \label{algo_line:stale_local_queues}
compute $dist(P, Q)$\;
add $P$ into $S$\;
\While{true} { \label{algo_line:stale_global_step_begin}
    divide all unchecked vertices from $S$ into $LS$\; \label{algo_line:stale_dispatching}
    \If{all $LS$ are empty} { \label{algo_line:stale_complete_start}
        break\;
     } \label{algo_line:stale_complete_end}
    \ForEach{worker $t$ out of $M$ \textbf{in parallel}} {
        \While{$LS[t]$ contains unchecked vertices {\bf and} $doMerge$ is \emph{false}} { \label{algo_line:stale_substate_start}
            vertex $v \gets$ the first unchecked vertex in $LS[t]$\;
            mark $v$ as checked\;
            \ForEach{neighbor $u$ of $v$ in $G$} { \label{algo_line:stale_neighbor_start}
                \If{$u$ is not visited} {\label{algo_line:stale_check_map}
                    mark $u$ as visited\; \label{algo_line:stale_update_map}
                    compute $dist(u, Q)$\;
                    add $u$ into $LS[t]$\;
                }
            } \label{algo_line:stale_neighbor_end}
            \textbf{if} $LS[t]$.size() $> L$, \textbf{then} $LS[t]$.resize($L$)\;
            \If{$t$ is the \emph{checker} {\bf and} CheckMetrics() returns \emph{true}} { \label{algo_line:checker_true}
                $doMerge \gets$ true\;
                assign the next checker in round-robin way\;
            }
       } \label{algo_line:stale_substate_end}
    }
    merge $LS$ into $S$\; \label{algo_line:stale_merge}
    \textbf{if} $S$.size() $> L$, \textbf{then} $S$.resize($L$)\;
    \If{$M < T$} {$M \gets 2M$}
} \label{algo_line:stale_global_step_end}
\Return the first $K$ vertices in $S$\;
\end{algorithm}

%% file: text/eval.tex
\section{Evaluation}\label{sec:eval}

\begin{figure*}[ht]
    \centering
    \includegraphics[width=\textwidth]{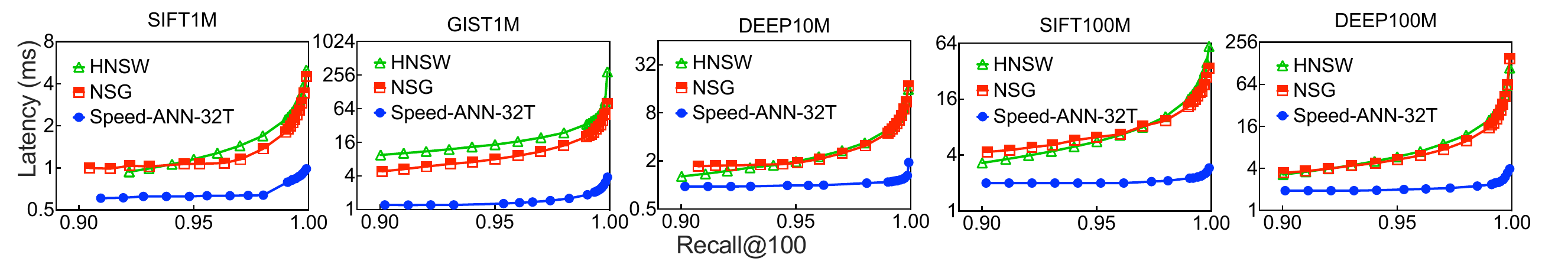}
    \caption[latency compared with baselines]{Latency (ms) comparison among \Hammer, NSG, and HNSW on five datasets. \textmd{\Hammer use 32 threads.}}
    \label{fig:eva_runtime_KNL}
    \vspace{-1em}
\end{figure*}

% This section evaluates \Hammer by comparing it with three state-of-the-art graph-based ANN searching approaches, NSG~\cite{}, HNSW~\cite{}, and a parallel version of NSG, and demonstrates that \Hammer outperforms them by up to  $XX\times$ speedup. Particularly, this section also demonstrates \Hammer's superior search efficiency on a large dataset with billions of points (achieving $XX$ latency for $XX$ search recall), and compares it with a state-of-the-art GPU implementation ($XXX$), proving \Hammer even outperforms it by $XX\times$ speedup with $XX$ CPU cores.     

This evaluation proves that  \Hammer can significantly reduce the ANN search latency with the proposed effective parallel optimizations. 

\PunchStarter{Evaluation Objectives.} This evaluation targets five specific evaluation objectives: 
(1) \textbf{latency}---demonstrating that \Hammer outperforms existing ANN search algorithms (NSG~\cite{fu2019fast}, HNSW~\cite{malkov2020efficient}, and a parallel version of NSG) by up to $76.6\times$ speedup in terms of the latency without any precision compromise; 
(2) \textbf{scalability}---confirming that \Hammer scales well on modern multi-core CPU architectures with up to 64-cores; 
(3) \textbf{optimization effects}---studying the performance effect of our key optimizations (parallel neighbor expansion, staged search, redundant-expansion aware synchronization, and cache friendly neighbor grouping) on overall latency, distance computations, synchronization overhead, etc; 
(4) \textbf{portability}---proving \Hammer has good portability by evaluating it on other multi-core CPU architectures; 
% (4) {\bf portability}---proving \Hammer has good portability by evaluating it on other multi-core CPU architectures; 
(5) \textbf{practicability}---showing that \Hammer is practical, applicable to extremely large datasets (e.g., \texttt{bigann}) with billions of points and outperforming an existing GPU implementation (i.e. Faiss) by up to $6.0\times$ speedup with $32$ CPU cores.

\begin{table}[t]
% \small
%\vspace{1em}
\caption{Characterization of  datasets. \textmd{ {\tt Dim.} denotes the dimension of the feature vector of each point, {\tt \#base} denotes the number of points, and {\tt \#queries} denotes the number of queries.}}\label{tab:datasets}
\centering
\begin{tabular}{|c|r|r|r|}
    \hline
    Dataset  & Dim. & \#base & \#queries \\ 
    \hline
    \hline
    SIFT1M  &       128 &      1M &        10K \\
    GIST1M  &       960 &      1M &         1K \\
    DEEP10M  &        96 &     10M &        10K \\
    SIFT100M &       128 &    100M &        10K \\
    DEEP100M &        96 &    100M &        10K \\ \hline
\end{tabular}
\end{table}

\PunchStarter{Implementation.}
A natural question is if our implementations
% , including \Hammer and \TopM (\TopMShortName)
%, and \ScaleM (\ScaleMShortName), 
can leverage any existing graph libraries (e.g., Ligra~\cite{shun2013ligra}); however, it turns out this is very difficult due to multiple reasons:
First, ANN algorithms do not pass messages between vertices. The computation only happens between a vertex and the query point.
Second, ANN algorithms need to do computation with vector values.
Third, ANN algorithms need to keep output results sorted. This requires extra efforts to maintain the results especially after synchronization between workers.
Fourth, existing libraries' optimization techniques for general graph processing are usually not suitable for ANN algorithms. For example, Ligra~\cite{shun2013ligra} can switch between push and pull modes 
% \minjia{Ligra needs a citation.}
according to the number of active vertices. However, in ANN algorithms, the number of active vertices is capped by the expected output number of nearest neighbors, making the switching never happen.
Besides, \Hammer runs in a semi-synchronous pattern with delayed synchronization among workers, which is different from the BSP model~\cite{valiant1990bridging} with strict synchronization after every parallel step.
Therefore, we have our high-performance implementation of those algorithms without using existing graph processing libraries. Our proposed ANN algorithms
%and \ScaleMShortName)
are written in C++ compiled by Intel C++ Compiler 2021.4.0 with ``\verb|-O3|'' option. We use OpenMP 5.0 to handle the intra-query parallelism.

%\paragraph{Evaluation settings}
\PunchStarter{Platform and Settings.}
Unless otherwise specified, all major experiments are conducted on Intel Xeon Phi 7210 (1.30 GHz) with 64 cores and 109 GB DRAM (\emph{KNL} for short).
\Hammer sets the average update position ratio as $0.8$ for SIFT1M, GIST1M, and SIFT100M, and $0.9$ for DEEP10M and DEEP100M.
%\TopMShortName set $M = 64$ for SIFT1M, DEEP10M, and SIFT100M, and $128$ for GIST1M and DEEP100M, respectively.
% \Hammer uses static synchronization settings in the major results to show its best latency performance and then compares to its adaptive version in later experiments.

%\paragraph{Datasets}
\PunchStarter{Datasets.}
This evaluation uses five datasets that are characterized in Table~\ref{tab:datasets}.
SIFT1M and GIST1M are from the datasets\footnote{\url{http://corpus-texmex.irisa.fr/}} introduced by J\'egou et al.~\cite{jegou2010product};
SIFT100M is sampled from the SIFT1B (\texttt{bigann}) introduced by J\'egou et al.~\cite{jegou2011searching};
DEEP10M and DEEP100M are sampled from DEEP1B\footnote{\url{https://sites.skoltech.ru/compvision/noimi/}} which is released by Babenko and Lempitsky~\cite{babenko2016efficient}.
These are common datasets for ANN algorithms evaluation~\cite{fu2019fast}.

%\paragraph{Baselines}
\PunchStarter{Baselines.}
\Hammer is compared with two state-of-the-art sequential ANN search implementations, NSG\footnote{\url{https://github.com/ZJULearning/nsg}}~\cite{fu2019fast} and HNSW\footnote{\url{https://github.com/nmslib/hnswlib}}~\cite{malkov2020efficient}. 
NSG employs a search algorithm called \SeqFullName, and HNSW uses its own best-first search algorithm corresponding to its hierarchical index.
%We use their default parameter settings for index building and searching.
The hyperparameters used for building their indices are set as default values as long as the authors provided them. Otherwise, several values are tested and the best performance is reported.
For NSG, we use its optimized version of searching for SIFT1M, GIST1M, and DEEP10M, and its normal version for SIFT100M and DEEP100M because of memory limit.
We also implement a Na\"ive Parallel NSG that parallelizes neighbor visiting during expansion.
%We use the implementation of \SeqFullName from NSG\footnote{\url{https://github.com/ZJULearning/nsg}} as the baseline. 
%For SIFT1M, GIST1M, and DEEP10M, we use NSG's optimized version of searching, while for SIFT100M and DEEP100M, we use its normal version because of memory limit.

\begin{figure}
%\vspace{-1em}
    \centering
    \includegraphics[width=0.45\textwidth]{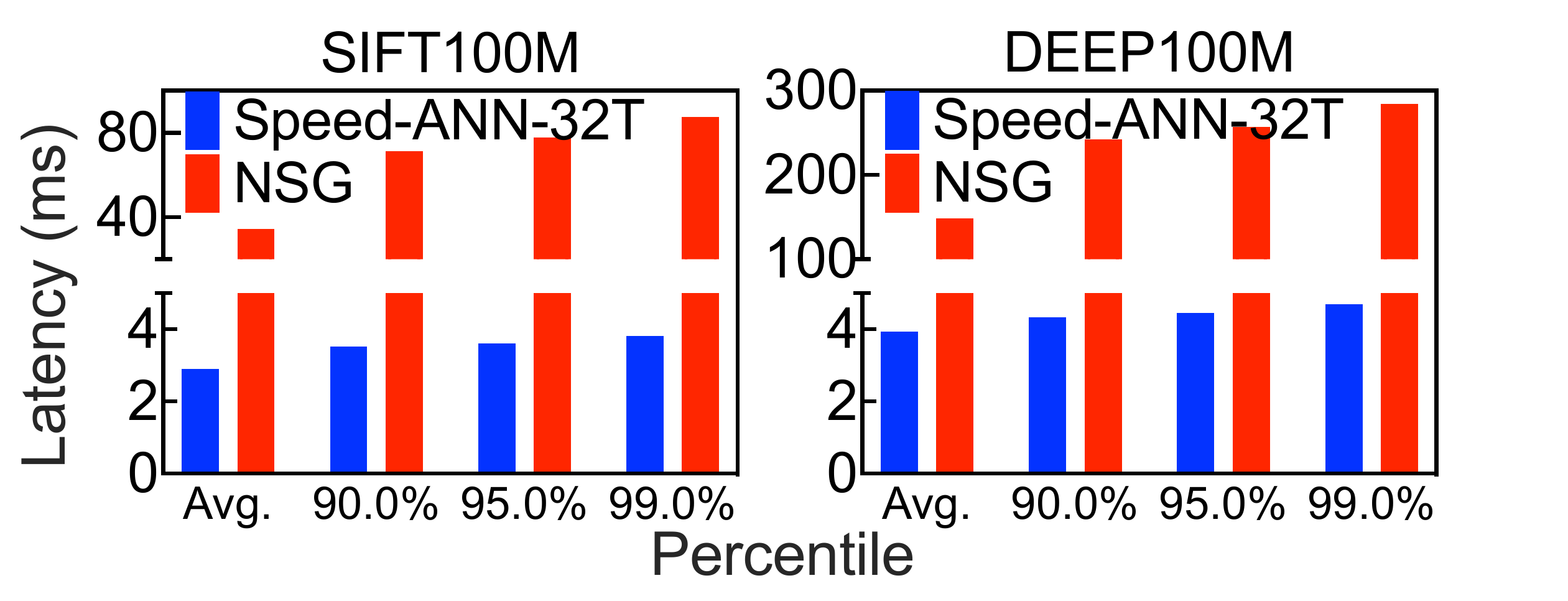}
    \caption[Percentile Latency]{Percentile latency of \Hammer\& NSG. Recall: 0.999.
    }
    \label{fig:eva_percentile_latency_KNL}
% \minjia{The pdf looks broken before Section 5.1. Using vspace with caution.}
    \vspace{3em}
\end{figure}

\subsection{Search Latency Results}
%\subsection{Overall Performance}

Figure~\ref{fig:eva_runtime_KNL} compares the latency of \Hammer, NSG, and HNSW.
\Hammer uses 32 threads while NSG and HNSW are sequential approaches.
The \emph{query latency} is the average latency of all queries, i.e., it equals the total searching time divided by the number of queries.
All methods search the 100 nearest neighbors for every query (i.e. $K = 100$).
The measure \emph{Recall@100} is calculated according to Formula~\ref{formula:recall} with $K = 100$, which means the ratio of ground-truth nearest neighbors in searching results for each query.
The value of Recall@100 is the average of all queries. All recalls mentioned in this section are Recall@100 if not specified.

%\minjia{Ideally, we should have recall@1 results (K=1).}

Figure~\ref{fig:eva_runtime_KNL} shows that \Hammer outperforms NSG and HNSW on all five datasets.
\Hammer's latency advantage increases with the growth of recall requirement, and it performs significantly better for high recall cases (e.g., from 0.995 to 0.999).
%When using 32 threads, \Hammer sometimes can hardly reach low recall (e.g. the short curve for SIFT100M) because such a wide speculative width 32 tends to achieve a high recall.
%Figure~\ref{fig:eva_speedup_PSS_to_NSG_HNSW} shows \Hammer's speedup over NSG and HNSW when \Hammer uses 32 threads. 
For the cases of Recall@100 (R@100) being 0.9, 0.99, and 0.999, on all five datasets,
\Hammer achieves $2.1\times$, $5.2\times$, and $13.0\times$ geometric mean speedup over NSG, 
and $2.1\times$, $6.7\times$, and $17.8\times$ over HNSW, respectively. 
As the recall becomes 0.999, \Hammer achieves up to $37.7\times$ speedup over NSG on DEEP100M, and up to $76.6\times$  speedup over HNSW on GIST1M.
\Hammer achieves significantly better performance for high recall situations mainly because of two reasons.
First, \Hammer's parallel neighbor expansion effectively reduces convergence steps (comparing with NSG) because it is not easily trapped at a local optimum and can explore a local region more quickly than a sequential search. This is particularly critical for a large graph (e.g., DEEP100M) to achieve high recall, where a query can more easily get stuck at a local optimum.
Second, \Hammer has better data locality from using aggregated L1/L2 cache provided by multiple threads, in contrast to sequential search where only private cache can be used.
Further profiling results are provided in Section~\ref{subsec:opt_effects}.

%\zhenfix{
\textbf{Impact on Tail Latency.}
For online inference, tail latency is as important, if not more, as the mean latency. To see if \Hammer provides steady speed-ups, we collect the 90th percentile (90\%tile), 95th percentile (95\%tile), and 99th percentile (99\%tile) latency from running NSG and \Hammer on SIFT100M and DEEP100M in Figure~\ref{fig:eva_percentile_latency_KNL}. The results show that 
while NSG's 99\%tile increases significantly by $154\%$ and $91\%$ for SIFT100M and DEEP100M, respectively, the \Hammer's 99\%tile increases only by $31\%$ and $19\%$ over its average for SIFT100M and DEEP100M, respectively. \Hammer leads to a relatively smaller increase in tail latency presumably because intra-query parallel search is particularly effective in reducing latency on long queries.
%}

\subsection{Scalability Results} \label{subsec:eval_scalability}
\begin{figure}
    \centering
    \includegraphics[width=0.45\textwidth]{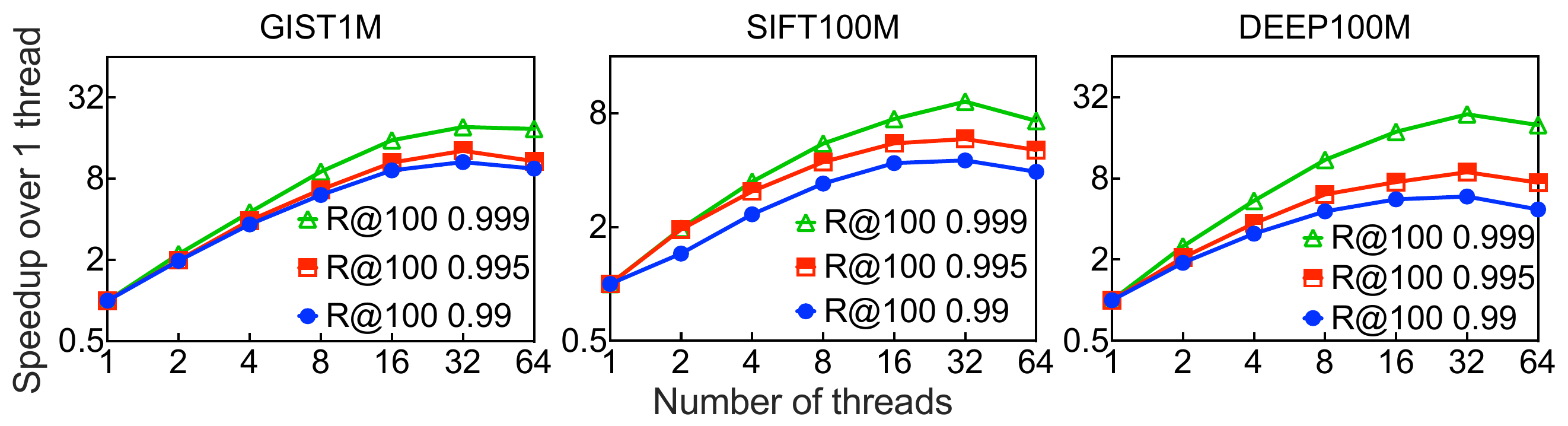}
    \caption[Scalability of PSS]{Speedup of \Hammer over 1 thread on five datasets.}
    \label{fig:eva_speedup_KNL}
    \vspace{-2em}
\end{figure}

\begin{figure}
    \centering
    \includegraphics[width=0.45\textwidth]{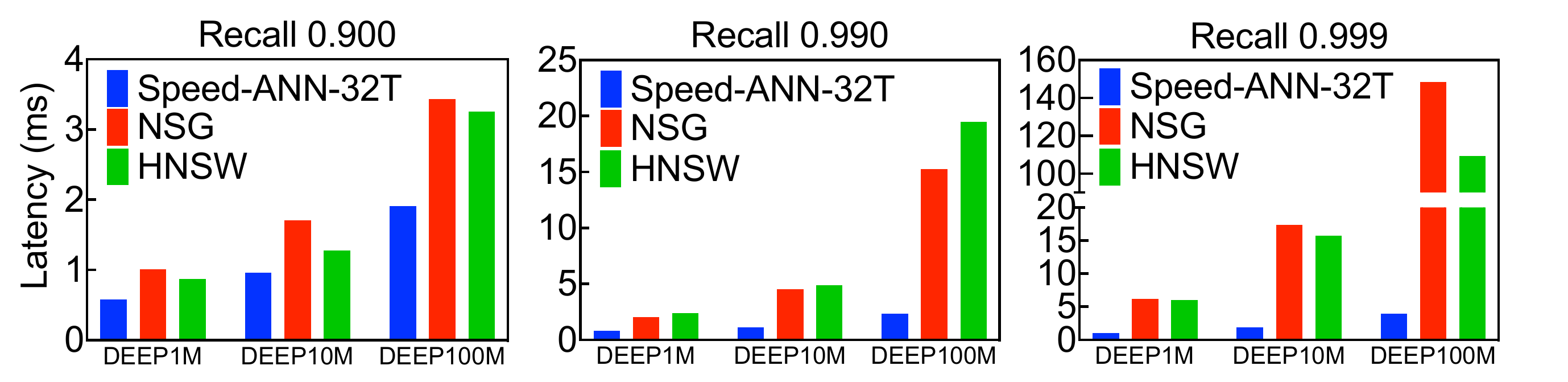}
    \caption[Scalability of Data Sizes]{Scalability with varied graph sizes for \Hammer, NSG, and HNSW on DEEP1M, DEEP10M, and DEEP100M. 
        \textmd{\Hammer uses 32 threads.}}
    \label{fig:eva_datasets_size_scale}
\end{figure}

%\subsubsection{Scalability with the Number of Threads} 
\PunchStarter{Scaling with An Increasing Number of Threads.} 
Figure~\ref{fig:eva_speedup_KNL} reports the speedup of 1- to 64-thread \Hammer over 1-thread on three datasets for three selected recall (0.99, 0.995, and 0.999), respectively. 
It shows that this scalability increases as the target recall grows because of the increased distance computations that offers more parallelism opportunities. 
The geometric mean speedup of all datasets for the highest recall (0.999) is $9.6\times$,
% \minjia{What's the reason for using geometric mean? Should it be arithmetic mean instead?}
$11.1\times$, and $9.2\times$ for 16-, 32-, and 64-thread, respectively.
%
%which provide enough workload for multiple threads to show their computing capability. 
%
%\textcolor{red}{TODO: Rephrase from here...}
%
\Hammer only scales to 16 threads for SIFT1M because SIFT1M is too small without enough workload for more threads.
\Hammer demonstrates super-linear speedup (up to 16 threads) for 0.999 recall on GIST1M and DEEP100M. 
This phenomenon will be further analyzed in Section~\ref{subsec:opt_effects}.
\Hammer does not scale well for 64 threads due to various reasons. For datasets with high dimensional vectors (e.g. GIST1M), 32-thread \Hammer has saturated memory bandwidth already. For others (e.g., SIFT1M, DEEP10M, and DEEP100M),  extra distance computations of too many unnecessary expansions gradually dominate overall execution. 
\PunchStarter{Scaling with An Increase of the Graph Sizes.}
%\TODO{NSG, HNSW, and \Hammer on DEEP1M/DEEP10M/DEEP100M.}
Our experiments also evaluate the scalability with varied dataset sizes (DEEP1M, DEEP10M, and DEEP100M) for \Hammer, NSG, and HNSW, respectively.
Figure~\ref{fig:eva_datasets_size_scale} reports the latency results of \Hammer, NSG, and HNSW for the recall of 0.9, 0.99, and 0.999, in which \Hammer uses 32 threads.
\Hammer constantly outperforms NSG and HNSW, and the heavier workload, the better performance \Hammer shows.
More specifically, with the growth of dataset size, the speedup of \Hammer over NSG and HNSW increases. For example, when the recall is 0.999, the speedup of \Hammer over NSG grows from $5.9x$ to $27.8x$ when the dataset size changes from 1M to 100M. This trend becomes increasingly obvious with the growth of the recall.
The results reflect that \Hammer is particularly effective and offers more speedups than existing search methods for larger graphs. 
% searching using multiple threads is more likely to find the nearest neighbors than sequential searching.

%\subsection{Study of Optimization Effects}
\subsection{Analysis Results}
\label{subsec:opt_effects}
% \TODO{need revision}

This section performs a series of experiments to show where \Hammer's improvements come from. It first compares \Hammer's performance with several alternative parallel search schemes. 
(i) NSG-32T: This config extends NSG with parallel neighbor expansion only (e.g., M=1). 
(ii) {\Hammer-\emph{NoStaged}}: This config is \Hammer but without using the staged search process.
(iii) {\Hammer-\emph{NoSync}}: This config performs parallel neighbor expansion but never synchronizes among workers until the very end. 
%{\tt \Hammer-Static} uses a static schedule where all threads bulk-synchronize in a given step limit.
(iv) {\Hammer-\emph{Exhaust}}: This config uses an exhaustive search to preprocess the dataset and obtain the proper synchronization settings. It should have the best latency performance, although requiring more than ten hours of tuning for the given dataset.
% \minjia{Is it true that it takes >10 hours to tune a dataset?}
(v) {\Hammer-\emph{Adaptive}}: This is the configuration described in Section~\ref{sec:design}, which adopts redundant-expansion aware synchronization.
%{\tt \TopMShortName-32T} accelerates the convergence with multi-path parallelism.
%{\tt \ScaleMShortName-32T} restricts unnecessary distance computation.
%{\tt \Hammer-32T} retains the benefit of parallelism and also reduces synchronization and unnecessary distance computation.

\begin{figure}
    
    \begin{subfigure}[t]{0.23\textwidth}
        \centering
        \includegraphics[height=1.in]{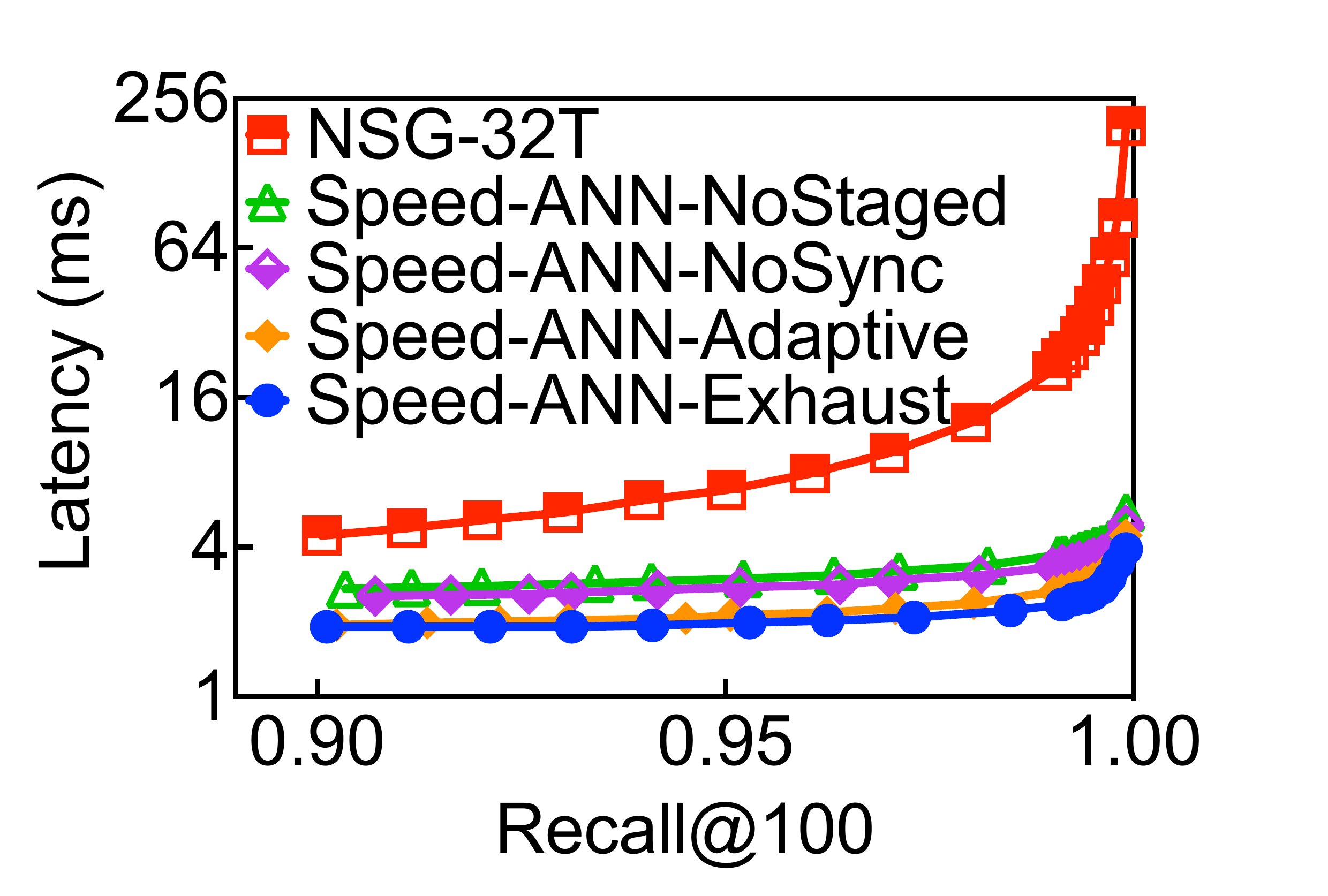}
        %    \caption[latency of parallel methods]{Performance of latency (ms.) for \Hammer, Na\"ive Parallel NSG, \TopM (\TopMShortName), and \ScaleM (\ScaleShortName) on five datasets using 32 threads.}
        \caption{Latency (ms)}
        \label{fig:eva_improvement_KNL}
    \end{subfigure}
    \hfill
    \begin{subfigure}[t]{0.23\textwidth}
        \centering
        \includegraphics[height=1.in]{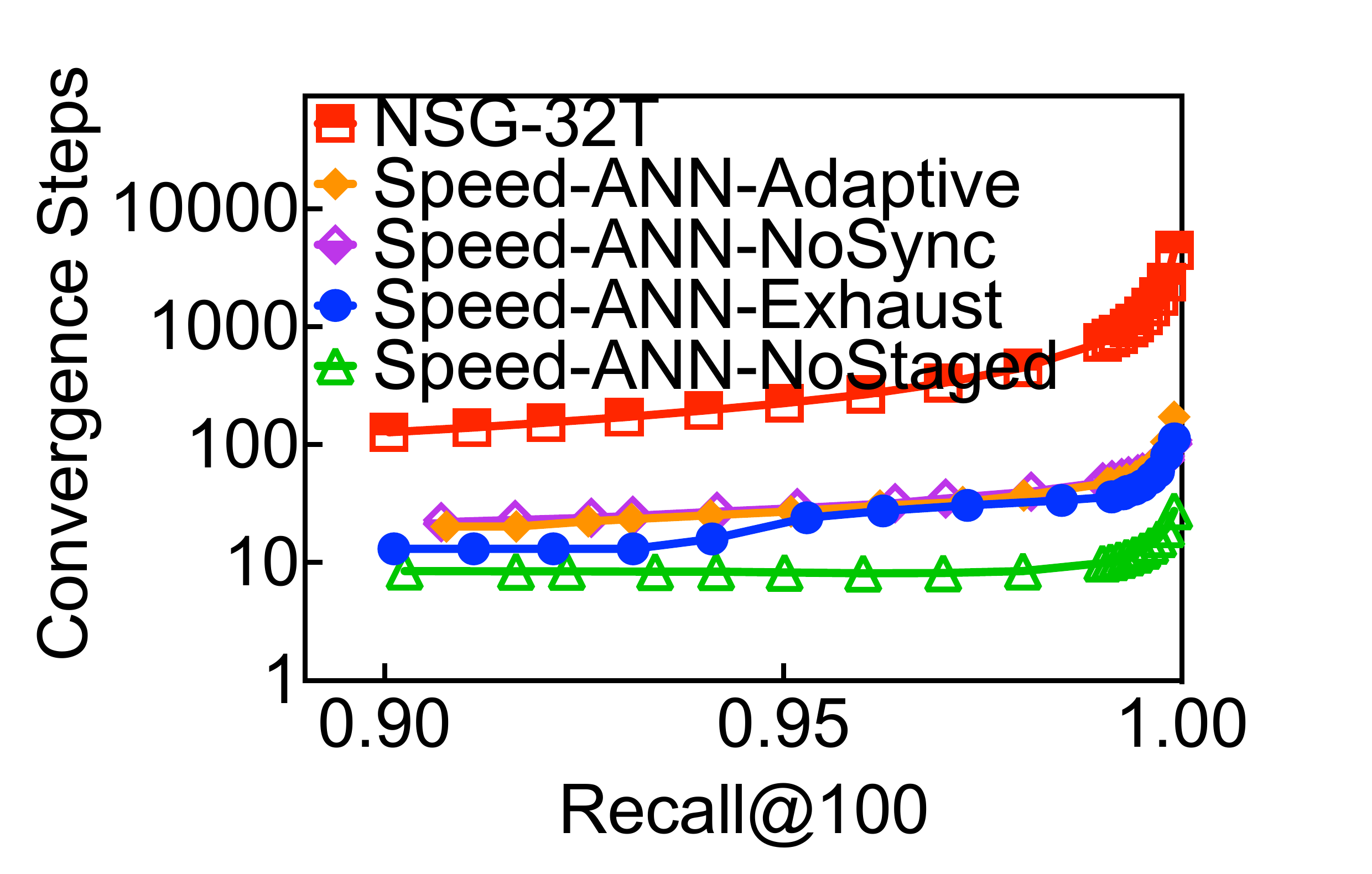}
        %    \caption[Convergence]{Convergence steps of \Hammer, Na\"ive Parallel NSG, \TopM (\TopMShortName), and \ScaleM (\ScaleShortName) on five datasets using 32 threads.}
        \caption{Convergence steps}
        \label{fig:eva_convergence_KNL}
    \end{subfigure}
    
    \begin{subfigure}[t]{0.23\textwidth}
        \centering
        \includegraphics[height=1.1in]{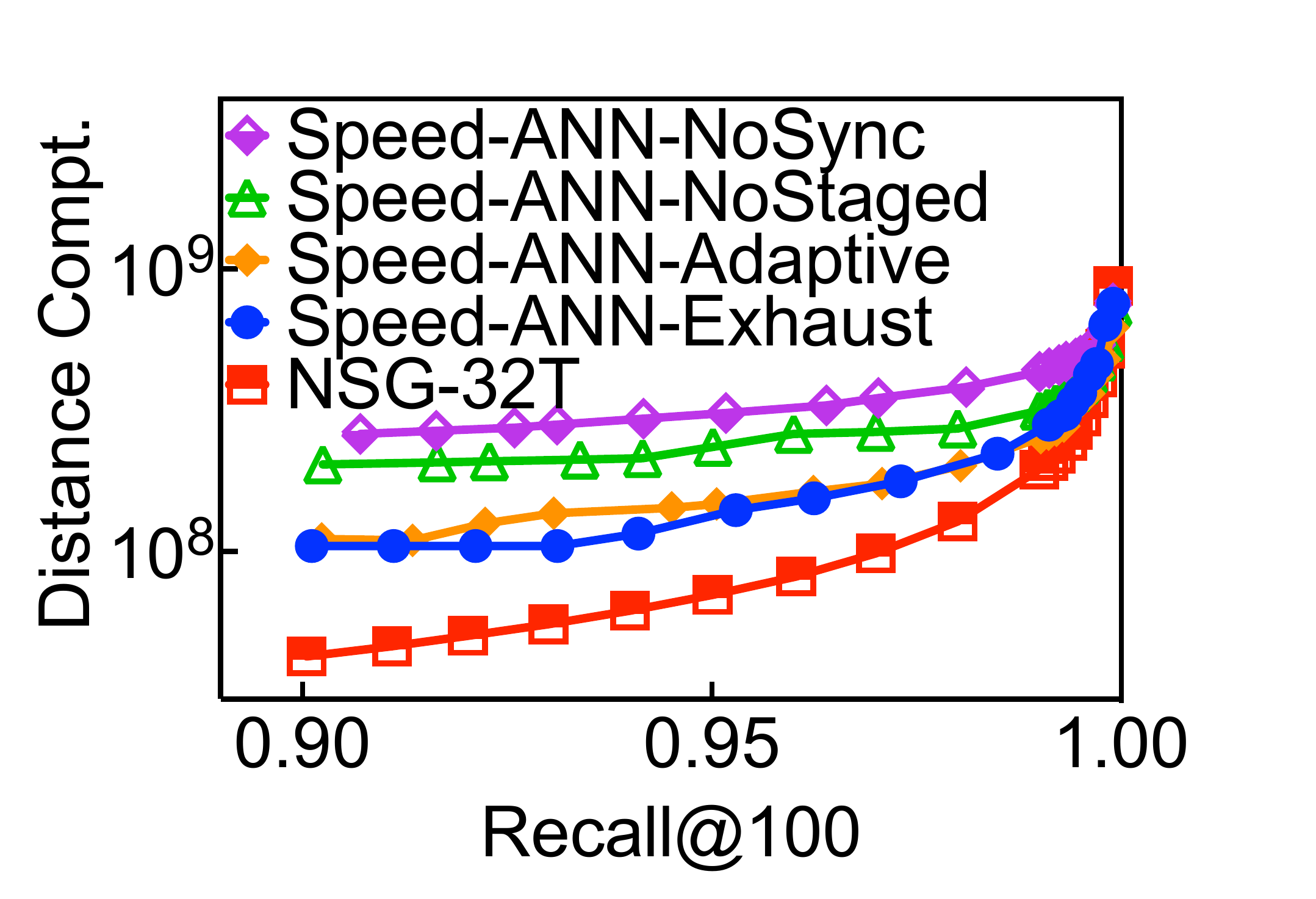}
        %    \caption[Computation]{Distance computation of \Hammer, Na\"ive Parallel NSG, \TopM (\TopMShortName), and \ScaleM (\ScaleShortName) on five datasets using 32 threads.}
        \caption{Distance computation}
        \label{fig:eva_computation_KNL}
    \end{subfigure}
    \hfill
    \begin{subfigure}[t]{0.23\textwidth}
        \centering
        \includegraphics[height=1.08in]{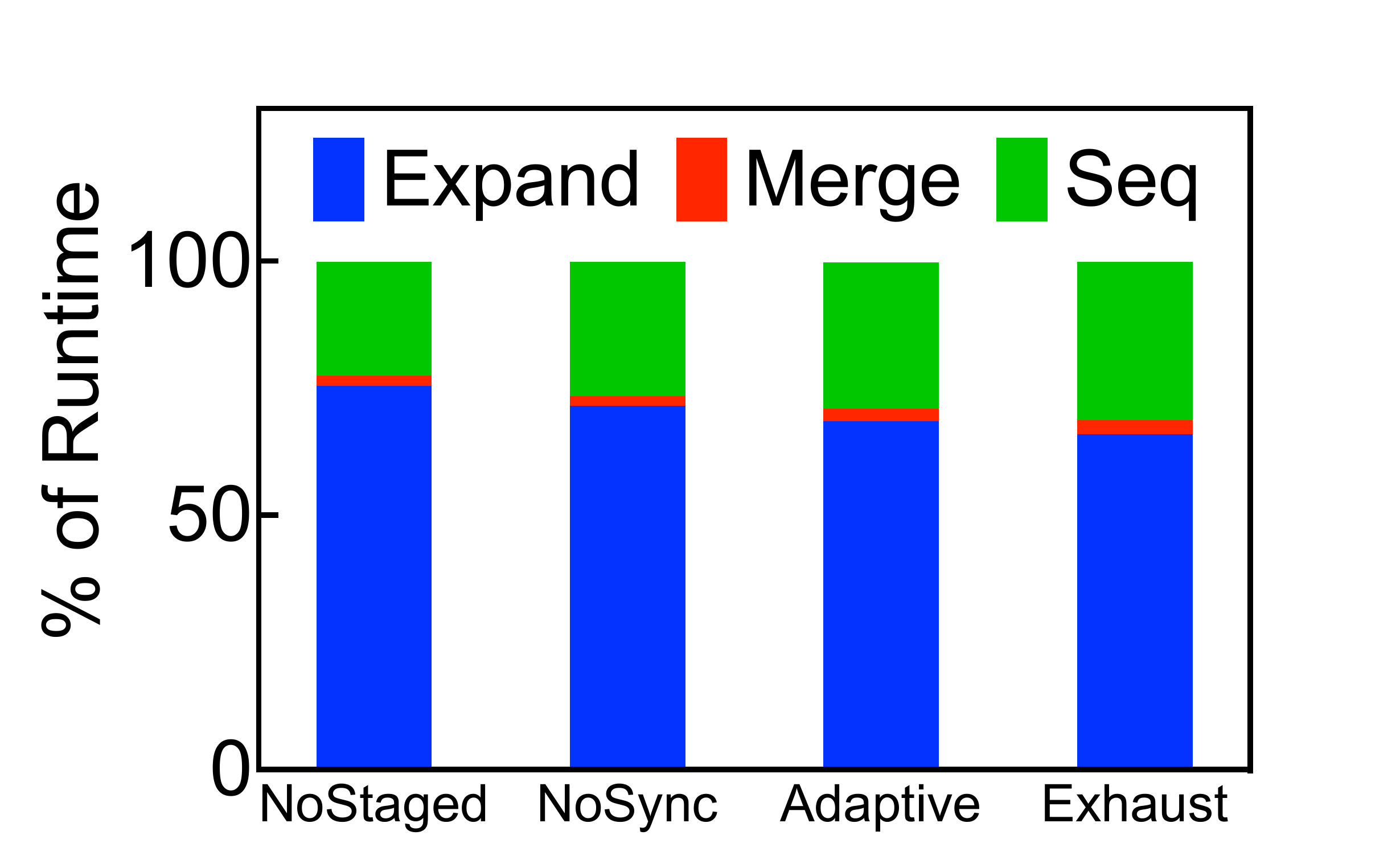}
        %    \caption[Breakdown]{Runtime breakdown of \Hammer, \TopM (\TopMShortName), and \ScaleM (\ScaleShortName) on five datasets using 32 threads for recall 0.999.
            %        \textmd{\emph{Expand} is when workers expand their candidates simultaneously. 
                %            \emph{Merge} is when workers merge their candidates.
                %            And \emph{Seq} is other sequential procedure.}}
        \caption{Runtime breakdown}
        \label{fig:eva_breakdown}
    \end{subfigure}
    \caption[Study of Synchronization]{Synchronization study w/ 32 threads on DEEP100M.}
    \label{fig:collect_perform_analysis}
\end{figure}

For this comparison, we report results on DEEP100M dataset with 32 threads in Figure~\ref{fig:collect_perform_analysis}. Other datasets and threads show the same trend, thus we omit them due to the space constraint. 
% Figure~\ref{fig:collect_perform_analysis} shows the comparison results.

%\noindent{\bf Latency Improvement. }
\noindent{\bf Effects on Latency. }
Figure~\ref{fig:eva_improvement_KNL} first reports the latency results of all five versions when we change recall from 0.90 to 1.00.
Compared with {NSG-32T}, 
{\Hammer-\emph{NoStaged}} has $4.9\times$ speedup on average for all recall cases, because of the convergence iterations reduction from parallel neighbor expansion.
{\Hammer-\emph{Exhaust}} has an extra $1.5\times$ speedup over {\Hammer-\emph{NoStaged}} mainly due to its reduction in synchronization optimization.
%convergence iterations reduction ({\tt \TopMShortName-32T}) brings $1.7\times$ speedup on average for all recall cases, 
%additional distance computation reduction ({\tt \ScaleMShortName-32T}) brings extra $1.5\times$ speedup, and additional synchronization optimization and the design of local search ({\tt \Hammer-32T}) brings extra $2.7\times$ speedup, respectively. 
{\Hammer-\emph{Exhaust}} achieves slightly better performance than {\Hammer-\emph{Adaptive}} (e.g., $1.1\times$ speedup). However, {\Hammer-\emph{Adaptive}} does not require the expensive offline tuning process as {\Hammer-\emph{Exhaust}}. 

%This section continues analyzing these optimizations with more profiling results.

% \noindent{\bf Convergence Iterations Reduction.}
\noindent{\bf Effects on Convergence Iterations.}
%\TODO{Update description according to the figure updated.}
Figure~\ref{fig:eva_convergence_KNL} profiles the convergence steps of the five parallel methods. Each point is averaged from all queries.
{NSG-32T} results in the most steps of convergence; while
{\Hammer-\emph{NoStaged}} results in the fewest.
All three versions of {\Hammer} result in comparable convergence steps to {\Hammer-\emph{NoStaged}} that are much less than {NSG-32T}.
This is because {\Hammer-\emph{NoStaged}} employs a fixed and relatively large number of multiple paths throughout the searching, resulting in the most aggressive exploring.
%{\tt \ScaleMShortName-32T} adaptively changes the path number, and {\tt \Hammer-32T} adopts local best-first search, both restricting distance computations while slightly increasing convergence steps.     
{\Hammer-\emph{Adaptive}} and {\Hammer-\emph{Exhaust}} adopt staged search, which slightly increases the convergence steps but significantly reduces distance computations.
Meanwhile, {\Hammer-\emph{NoSync}} suffers more divergence compared to {\Hammer-\emph{Adaptive}} and {\Hammer-\emph{Exhaust}}.

% \PunchStarter{Distance Computation Reduction.}
\PunchStarter{Effects on Distance Computation.}
Figure~\ref{fig:eva_computation_KNL} profiles the number of distance computations for those five methods.
{\Hammer-\emph{NoStaged}} with a fixed value of $M=32$ leads to more distance computations than {NSG-32T}, {\Hammer-\emph{Exhaust}}, and {\Hammer-\emph{Adaptive}} to achieve the same recall (especially for low recall cases).
% \minjia{\tt causes the above to break the margin. Please fix it.}
While completely removing synchronization, {\Hammer-\emph{NoSync}} has the most distance computations than others.
However, as shown in Figure~\ref{fig:eva_improvement_KNL}, it still achieves lower latency than {\Hammer-\emph{NoStaged}} because synchronization overhead can dominate the total search time when the number of parallel workers is large.
%Both {\tt \ScaleMShortName-32T} and {\tt \Hammer-32T} can effectively restrict distance computations. 
%Comparing with {\tt \ScaleMShortName-32T}, {\tt \Hammer-32T} incurs more distance computations because of the loosely coupled parallel design that reduces the synchronization (particularly merge) overhead.
% In the high recall regime (e.g., 0.999), all \Hammer based variants result in fewer distance computations than {\tt NSG-32T}, indicating that our proposed is particularly effective for high recalls (on large graphs).

% \PunchStarter{Synchronization Overhead Study.}
\PunchStarter{Effects on Synchronization Overhead.}
Figure~\ref{fig:eva_breakdown} reports the execution time breakdown of our four approaches.
It splits the whole execution time into three parts: Expanding part (\emph{Expand}), Merging part (\emph{Merge}), and Sequential part (\emph{Seq}).
\emph{Expand} denotes the parallel phase of a query that workers expand their unchecked candidates. It consists of computing distances and inserting visited neighbors into their queues.
\emph{Merge} denotes the phase that workers merge their local queues into a global queue after they complete expanding. It reflects the major  synchronization overhead.
Other sequential execution of a search is included in \emph{Seq}.
All results are for recall 0.999.
Figure~\ref{fig:eva_breakdown} shows that redundant-expansion aware synchronization strategy effectively mitigates the synchronization overhead, allowing {\Hammer-\emph{Adaptive}} to achieve a similar portion of synchronization overhead as {\Hammer-\emph{Exhaust}}.
% does.
% {\Hammer-\emph{Exhaust}} has the smallest portion of \emph{Expand} thanks to its tuned settings of parallelism.
% %%%%%%%%%%%%%%%%%%%%%%%%%%%%%
% %% backup
% Figure~\ref{fig:eva_breakdown} validates that {\tt \Hammer}'s \emph{Merge} is much more efficient than {\tt \TopMShortName}' due to its loose synchronization, and hence its other parts (\emph{Expand} and \emph{Seq}) take up more execution percentages.
% Among three versions of {\tt \Hammer}, {\tt \Hammer-\emph{NoSync}} has the smallest portion of \emph{Merge} than both {\tt \Hammer-\emph{Exhaust}} and {\tt \Hammer-\emph{Adaptive}}.
% %% end backup
% %%%%%%%%%%%%%%%%%%%%%%%%%%%%%

\begin{figure}
    \begin{minipage}[t]{0.23\textwidth}
        \centering
        \includegraphics[height=0.95in]{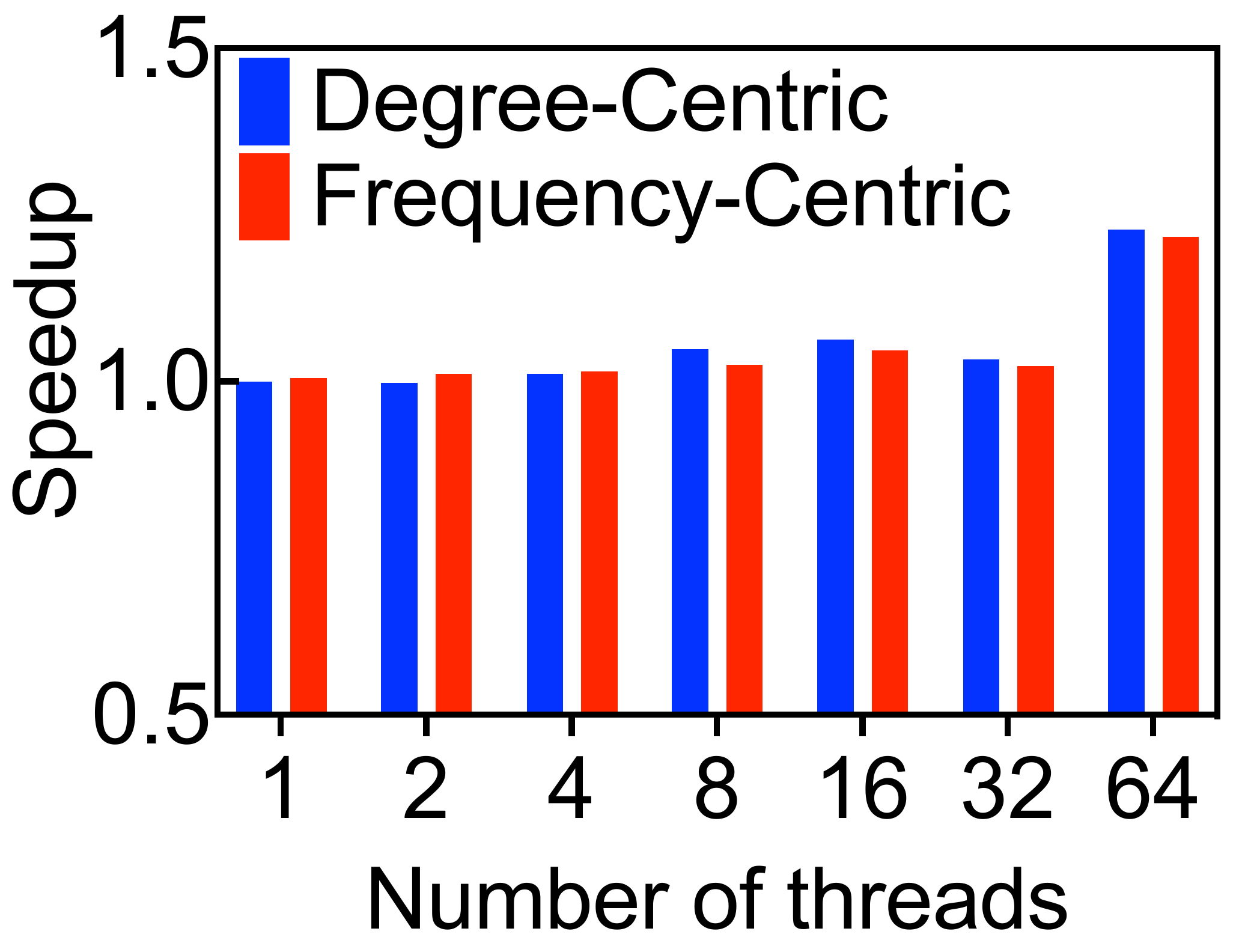}
        \caption[Reorder on KNL]{Speedup of \Hammer's neighbor grouping on DEEP100M for recall 0.999. 
            %            \bin{TODO: change figure order}
        }
        \label{fig:eva_reorder_KNL}
%        \vspace{-2em}
    \end{minipage}
    \hfill
    \begin{minipage}[t]{0.23\textwidth}
        \centering
        \includegraphics[height=1.in]{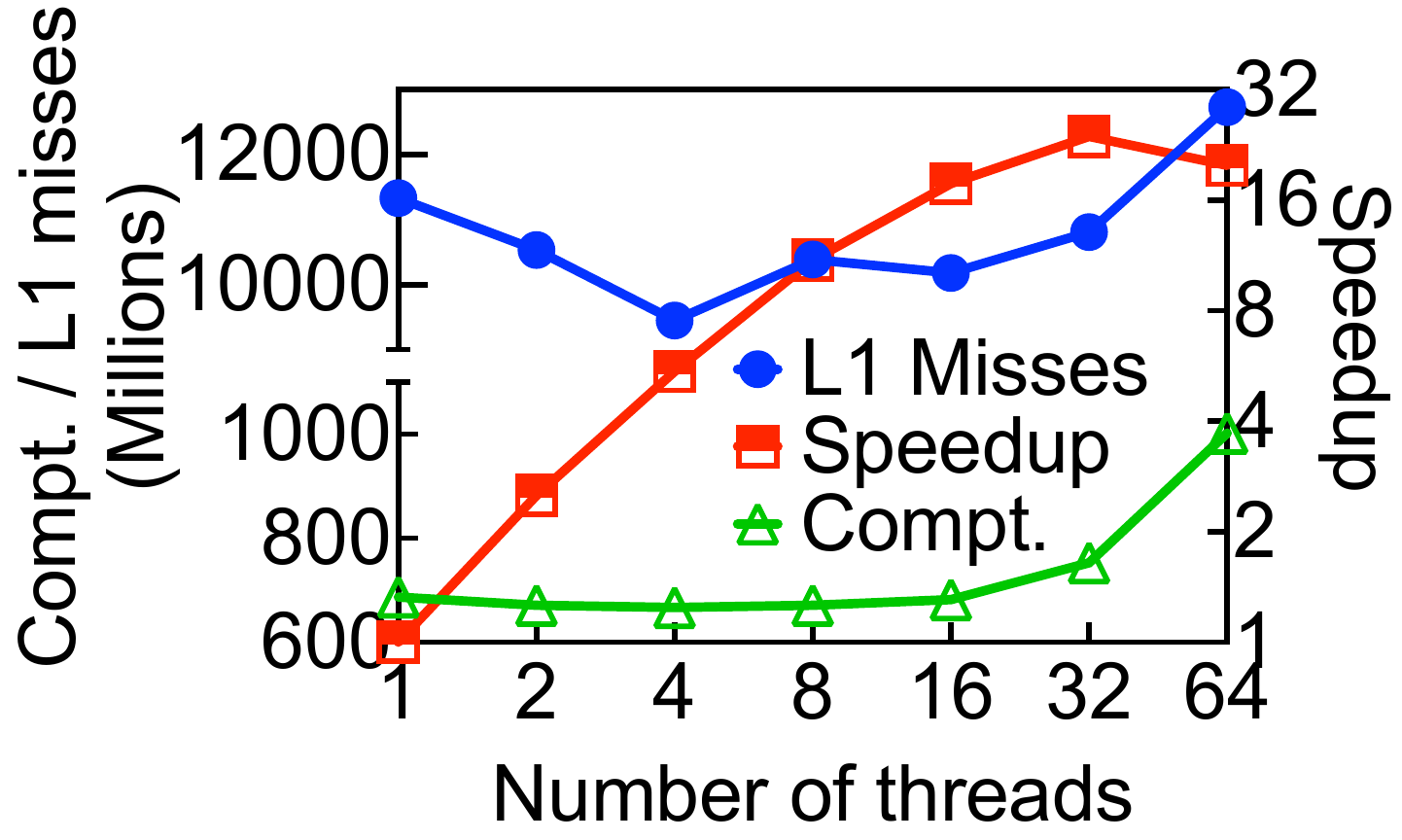}
        %    \caption[L1 misses, computation, and speedup]{Illustration about how L1 misses and distance computation (compt.) influence \Hammer's speedup over its 1 thread latency.}
        \caption[L1 misses, computation, and speedup]{\Hammer's L1 misses, speedup over 1-thread, and distance computation w/ recall 0.999 on DEEP100M.}
        \label{fig:eva_cache_compt_speedup}
%        \vspace{-2em}
    \end{minipage}
\end{figure}

% \subsubsection{Breakdown analysis of additional optimizations.}

\PunchStarter{Effects of Neighbor Grouping.}
Our fully optimized {\Hammer-32T} also includes another optimization, i.e. neighbor grouping. Figure~\ref{fig:eva_reorder_KNL} shows that our two proposed strategies (\emph{degree-centric} and \emph{frequency-centric}) outperform no-grouping by up to $1.22\times$ and $1.21\times$ speedup, respectively, when we change the thread numbers from 1 to 64.  
This speedup mainly comes from the reduction of the last-level cache miss and TLB (translation lookaside buffer) cache miss. This profiling result is omitted due to the space constraint.

\PunchStarter{Super-linear Speedup Observation.}
Section~\ref{subsec:eval_scalability} shows that \Hammer results in an interesting super-linear speedup (up to 16 threads) for 0.999 recall on GIST1M and DEEP100M. 
% We explain the underlying reasons here because this is interesting.
Figure~\ref{fig:eva_cache_compt_speedup} reports three profiling results, distance computations, L1 cache misses, and performance speedup for DEEP100M when changing the thread numbers from 1 to 64. The left x-axis shows the first two profiling results while the right x-axis shows the last one. It shows that as we increase the number of threads, the L1 cache misses and distance computations first decrease and then increase. This causes the super-linear speedup for the cases whose thread numbers are less than 16.
Distance computation shows this trend because: on the one hand, parallel neighbor expansion helps avoid the search from being trapped by local minimal candidates and quickly pick up promising searching paths for more nearest neighbors; on the other hand, too many exploring threads cause unnecessary expansion of non-promising candidates, increasing distance computations. L1 cache miss shows this trend because multi-threads increase the total size of L1 cache.

\begin{figure}
    \begin{subfigure}[t]{0.23\textwidth}
        \centering
        \includegraphics[height=1.2in]{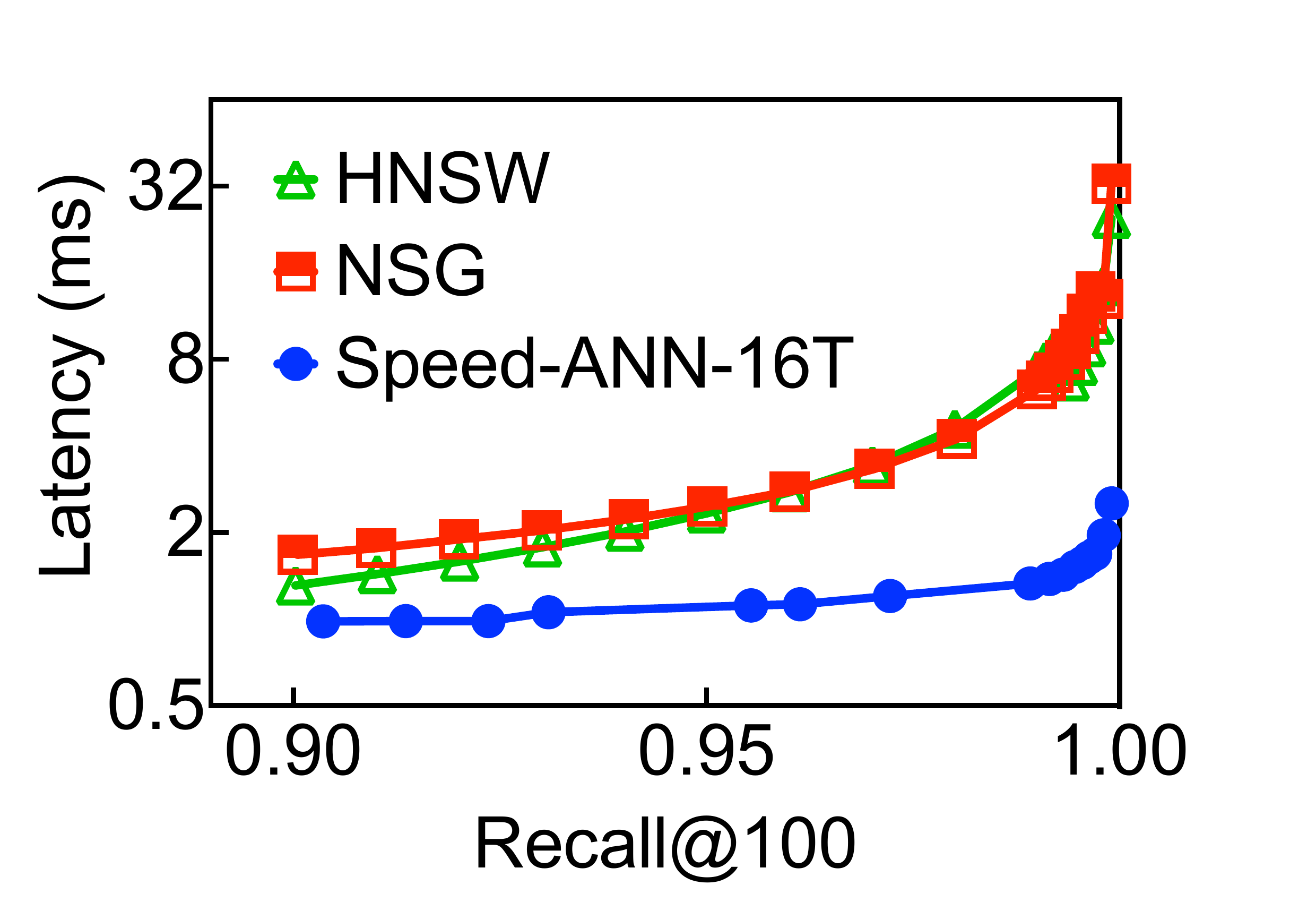}
        %    \caption[Skylake latency comapred with baselines]{Performance of latency (ms.) for \Hammer, NSG, and HNSW on five datasets on Skylake. \Hammer use 16 threads. }
        \caption{Latency (ms)}
        \label{fig:eva_runtime_Skylake}
    \end{subfigure}
    \hfill
    \begin{subfigure}[t]{0.23\textwidth}
        \centering
        \includegraphics[height=1.2in]{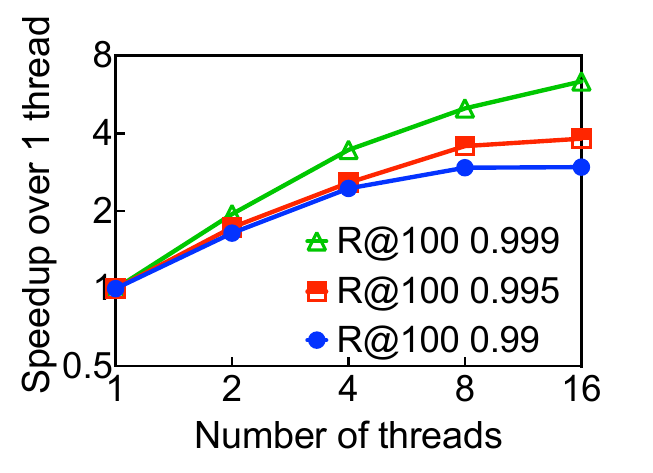}
        %    \caption[Skylake scalability of \Hammer]{Speedup of \Hammer over 1 thread on five datasets on Skylake.}
        \caption{Speedup}
        \label{fig:eva_speedup_Skylake}
    \end{subfigure}
    \caption{Portability study: DEEP100M on Skylake.}
    \vspace{-2em}
\end{figure}

\subsection{Portability Evaluation}
To evaluate the portability,  \Hammer is also tested on Intel Skylake architecture, Xeon Gold 6138 (2.0 GHz) with 20 cores and 187 GB DRAM ({\tt Skylake} for short). For the sake of space saving, only results on DEEP100M are presented as other datasets show the same trend.
Figure~\ref{fig:eva_runtime_Skylake} compares the latency of \Hammer, NSG, and HNSW, in which,
\Hammer uses 16 threads.
It shows a similar trend as previous, i.e., \Hammer outperforms NSG and HNSW for all recall.
For 0.9, 0.99, and 0.999 cases, 
\Hammer achieves $1.7\times$, $4.5\times$, and $12.9\times$ speedup over NSG, 
and $1.3\times$, $5.3\times$, and $9.7\times$ over HNSW, respectively. 
Figure~\ref{fig:eva_speedup_Skylake} evaluates \Hammer's scalability. 
Similarly, target recall 0.999 can achieve the best speedup over 1 thread, and 
speedup for 0.999 is $4.9\times$ and $6.3\times$ for 8 threads and 16 threads, respectively.

\begin{figure}
    \centering
    \includegraphics[width=0.45\textwidth]{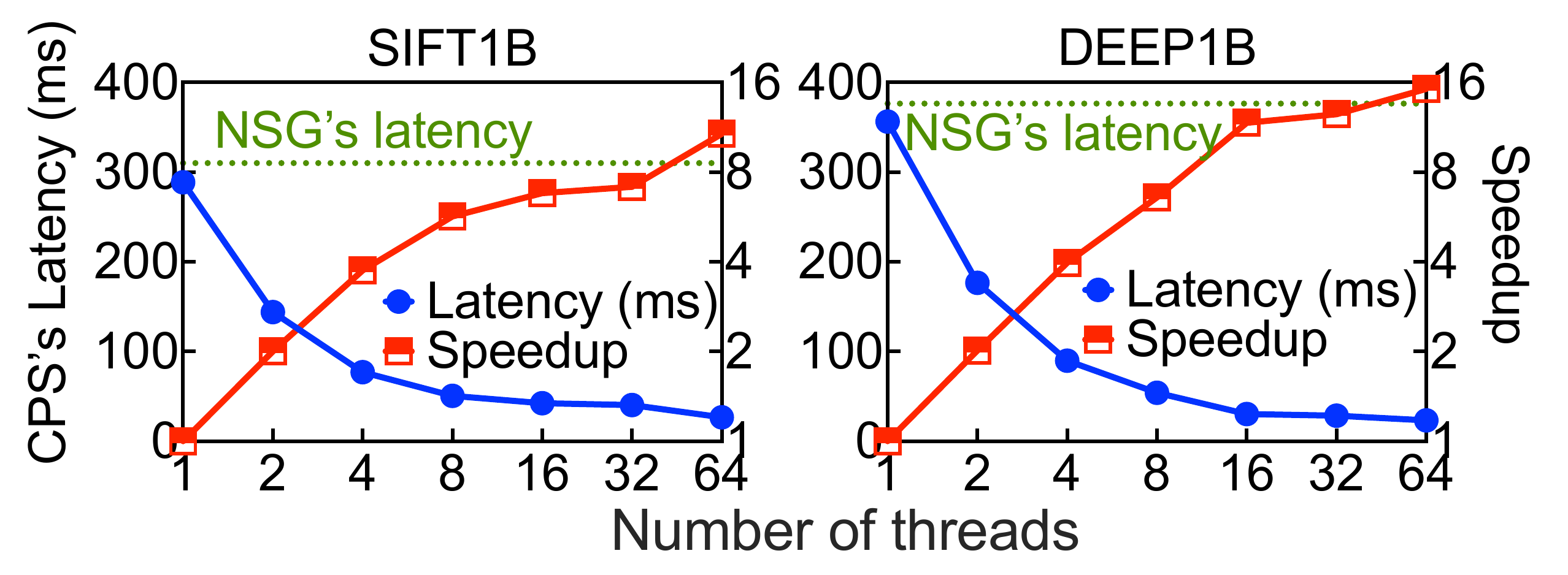}
    \caption[Latency for 1B datasets]{Performance comparison of \Hammer and NSG on SIFT1B (\texttt{bigann}) and DEEP1B. 
        \textmd{\Hammer's speedup is over its 1-thread. Recall is 0.9.}}
    \label{fig:eva_sift1b_deep1b_latency}
    % \vspace{-2em}
\end{figure}

\subsection{Practicality Evaluation}

This section evaluates \Hammer's practicality with two case studies: 1) evaluating it on very large datasets,  SIFT1B (\texttt{bigann}) and DEEP1B that contain over 1 billion data vectors; 2) comparing it with a state-of-the-art GPU implementation.

\PunchStarter{Billion-Scale Datasets.}
 This experiment is conducted on a particular machine with Xeon Gold 6254 (3.10 GHz) 72 cores and 1.5 TB memory because of the large memory requirement.
% It is worth mentioning that even 1.5 TB memory is not enough to build a 100-NN graph with one billion data vectors. Therefore, some sub-optimal parameters are chosen when generating the corresponding NSG index.
Figure~\ref{fig:eva_sift1b_deep1b_latency} compares the latency of \Hammer and NSG. \Hammer uses up to 64 threads, and the recall target is 0.9. 
When using 64 threads, \Hammer outperforms NSG with $11.5\times$ and $16.0\times$ speedup for SIFT1B and DEEP1B, respectively. As we increase the number of threads, \Hammer shows sub-linear speedup because of the well-known NUMA effect (this machine has 4 NUMA domains). These results indicate the effectiveness of our method in speeding up the search process on billion-scale datasets.  

%For \Hammer, when the number of threads is increasing, its speedup becomes sub-linear. It is probably because of the NUMA accesses as those cores are divided into four NUMA domains, and a single domain cannot hold the whole dataset.

\PunchStarter{Compare with a GPU Implementation.} We also compare \Hammer with a GPU-based large-scale ANN search algorithm~\cite{johnson2017billion} in Faiss library~\cite{faiss-code}. The GPU experiments are conducted on an NVIDIA Tesla P100 with CUDA 10.2. Faiss is set to have one query in every batch, because we focus on reducing the online query latency to meet stringent latency requirement.
Table~\ref{tab:gpu_latency} shows the latency comparison results on five datasets. \Hammer uses 32 threads on KNL.
For the SIFT100M and DEEP100M, Faiss-GPU complains of out-of-memory errors. 
For other datasets, \Hammer outperforms Faiss-GPU with $1.4\times$ to $6.0\times$ speedup and much better recall, which indicates that \Hammer can effectively achieve faster search speed than GPU-based search algorithms on CPUs, which are often much cheaper than GPUs. 
% \usepackage{multirow}
\begin{table}[]
\small
    \caption{Latency comparison of \Hammer and Faiss-GPU on five datasets. 
        \textmd{{\tt Lt.} means \emph{Latency}.
            {\tt OOM} means \emph{out of memory}. 
            Faiss-GPU's index format is IVFFLat.
            \Hammer uses 32 threads.}}
    \label{tab:gpu_latency}
    \begin{tabular}{|c|cc|c|c|}
        \hline
        \multirow{2}{*}{Datasets} & \multicolumn{2}{c|}{Faiss-GPU w/ IVFFlat}                             & \multicolumn{2}{c|}{\Hammer-32T on KNL}                                    \\ \cline{2-5} 
        & \multicolumn{1}{c|}{R@100}  & \multicolumn{1}{c|}{Lt. (ms.)} & \multicolumn{1}{c|}{R@100} & \multicolumn{1}{c|}{Lt. (ms.)} \\ 
        \hline \hline
        SIFT1M                    & \multicolumn{1}{c|}{0.52}          & 0.87                               & \textbf{0.91}                              & \textbf{0.61}                               \\
        GIST1M                    & \multicolumn{1}{c|}{0.36}          & 7.25                               & \textbf{0.90}                              & \textbf{1.21}                               \\
        DEEP10M                   & \multicolumn{1}{c|}{0.62}          & 5.79                               & \textbf{0.90}                              & \textbf{0.96}                               \\
        SIFT100M                  & \multicolumn{1}{c|}{OOM} & \multicolumn{1}{c|}{OOM}              & \textbf{0.90}                              & \textbf{2.00}                               \\
        DEEP100M                  & \multicolumn{1}{c|}{OOM} & \multicolumn{1}{c|}{OOM}              & \textbf{0.90}                              & \textbf{1.91}                               \\ \hline
    \end{tabular}
\end{table}

%\textcolor{red}{TODO: sift1b}

%% file: text/related.tex
%\clearpage
\section{Related Work}\label{sec:related}

%\TODO{shrink to summaries}

% There is significant work about approximate nearest neighbor search.
% In this work, we focus on parallelizing the search procedure on a given graph index.
%In contrast, many graph-based methods aim to build the graph index to represent data vectors.
%Meanwhile, there are non-graph-based methods, including hash-based methods, quantization-based methods, and tree-based methods.

This section describes prior efforts closely related to our work.

\noindent{\bf Graph-based ANN.} 
%\subsubsection{CPU}
%Graph-based indexing methods represent data points as vertices and their relation as edges.
Navigating Spreading-out Graph (NSG)~\cite{fu2019fast} is one of the state-of-the-art graph-based indexing methods. 
It is a close approximation of Monotonic Relative Neighborhood Graph (MRNG) that ensures a close-logarithmic search complexity with limited construction time.
NSG (and many other graph-based methods~\cite{dearholt1988monotonic,arya1993approximate,hajebi2011fast,jin2014fast,malkov2014approximate,malkov2020efficient,harwood2016fanng}, e.g., FANNG~\cite{harwood2016fanng}, NSW~\cite{malkov2014approximate}, and HNSW~\cite{malkov2020efficient}) rely on best-first search to process queries. 
%\minjia{Perhaps change "greedy search" to "best-first" search?}
\iffalse
Here are some examples.  
FANNG~\cite{harwood2016fanng} employs a backtrack search algorithm (i.e., a depth-first best-first search) that explores each edge of a candidate vertex and terminates once the total number of distance computations reaches a certain threshold (called {\em stop condition}).
NSW~\cite{malkov2014approximate} performs multiple best-first searches for a given query while all searches share the same result set. 
HNSW~\cite{malkov2020efficient} separates links into multiple layers according to their lengths, forming a hierarchical graph structure. A best-first search starts from the upper layer until reaching the bottom.
\fi
%
Other graph-based methods include~\cite{li2020approximate,zhang2019grip,subramanya32diskann,li2020improving,baranchuk2019learning,lin2019graph,prokhorenkova2020graph,deng2019pyramid,bashyam2020fast}.
\iffalse
For example, GRIP~\cite{zhang2019grip} and DiskANN~\cite{subramanya32diskann} consider the issue of storing indices in both DRAM and SSDs.
Pyramid~\cite{deng2019pyramid} and Bashyam et al.~\cite{bashyam2020fast} build distributed indices based on HNSW.
GGNN~\cite{groh2019ggnn} and SONG~\cite{zhao2020song} leverage GPUs for indices building and distance computing.
\fi
%
In contrast to these efforts that mostly focus on indexing building, our work {\em for the first time} unveils the real bottleneck of intra-query graph search, and significantly reduces search latency (particularly for billion-scale graphs) with multiple advanced architecture-aware parallel techniques.

\noindent{\bf Non-Graph based ANN Methods.} 
Hashing-based methods~\cite{indyk1998approximate,datar2004locality,andoni2006near,andoni2015practical} map data points into multiple buckets with a certain hash function such that the collision probability of nearby points is higher than the probability of others.
\iffalse
FALCONN~\cite{andoni2015practical} is a library based on locality-sensitive hashing~\cite{indyk1998approximate}.
\fi
%They can speedup the computations through a direct hash lookup. 
%LSH~\cite{indyk1998approximate} is a hashing-based method, which is basis of FALCONN library~\cite{andoni2015practical}.
%, although their recall is low when compared to other methods~\cite{jegou2010product}.
%
Quantization-based methods~\cite{jegou2008hamming,ge2013optimized,wu2017multiscale,wei2020analyticdb,wang2020deltapq} (e.g., IVF~\cite{jegou2010product}, and IMI~\cite{babenko2014inverted}) 
%group data into clusters such that a query is expected to be answered by searching in some clusters rather than the whole dataset.
%\minjia{Quantization-based methods 
compress vectors into short codes to reduce the number of bits needed to store and compute vectors. 
%It is orthogonal to how to group data.
Faiss~\cite{johnson2017billion} is implemented by Facebook with produce quantization (PQ) methods.
%AnalyticDB-V~\cite{wei2020analyticdb} uses HNSW at streaming level to deal with real-time data insertion and modification, while uses PQ to compress newly inserted vectors. 
%
Tree-based methods (e.g., KD-tree~\cite{silpa2008optimised} and R* tree~\cite{beckmann1990r}) hierarchically split the data space into lots of regions that correspond to the leaves of a tree structure, and only search a limited number of promising regions. 
%\minjia{Is DeltaPQ a quantization based method or tree-based method?}
Flann~\cite{muja2009fast} is a library based on KD-tree. 
%DeltaPQ~\cite{wang2020deltapq} compresses the quantization codes by storing their difference in a \emph{DeltaTree}.
%
Graph-based methods have been proved to outperform these non-graph-based methods by checking fewer data points to achieve the same recall~\cite{fu2019fast,ann-benchmark,li2020approximate,echihabi2019return}.
Another line of work that is closely related to \Hammer is to accelerate ANN search by varied accelerators, e.g., FPGA~\cite{fpga-ann} and GPU~\cite{johnson2017billion}.

%\begin{enumerate}
%    \item PVLDB-2020, DeltaPQ: Lossless Product Quantization Code Compression for High Dimensional Similarity Search. Compress the quantization codes by storing their difference in a \emph{DeltaTree}.
%    \item PVLDB-2020, AnalyticDB-V: a hybrid analytical engine towards query fusion for structured and unstructured data. Use HNSW at streaming level to deal with real-time data insertion and modification, while use product quantization to compress newly inserted vectors and rebuilds ANNS indices.
%    \item Methods introduced in NSG paper.
%    \item Methods introduced in PMLR-2019, Learning to Route in Similarity Graphs: tree-based, locality-sensitive hashing.
%    \item Methods introduced in SIGMOD-2020, PASE: PostgreSQL Ultra-High-Dimensional Approximate Nearest Neighbor Search Extension.
%\end{enumerate}

\noindent{\bf Parallel Graph Systems.} 
%Ligra. But cannot match the problem very well.
Many graph engines and frameworks have been developed in the past decade.
Some of them are shared-memory, focusing on processing in-memory datasets within a computation node, e.g., Galois~\cite{nguyen2013lightweight}, Ligra~\cite{shun2013ligra}, Polymer~\cite{zhang2015numa}, GraphGrind~\cite{sun2017graphgrind}, GraphIt~\cite{zhang2020optimizing}, and Graptor~\cite{vandierendonck2020graptor}.
Some are distributed systems, e.g., Pregel~\cite{malewicz2010pregel}, GraphLab~\cite{low2014graphlab}, and PowerGraph~\cite{joseph2012powergraph}.
Some efforts focus on out-of-core designs (e.g., GraphChi~\cite{aapo2012graphchi} and X-Stream~\cite{roy2013x}) and process large graphs with disk support.
Many graph frameworks are also on GPUs, such as CuSha~\cite{khorasani2014cusha}, Gunrock~\cite{wang2016gunrock}, GraphReduce~\cite{sengupta2015graphreduce}, and Graphie~\cite{han2017graphie}.
%
%\textcolor{red}{TODO: more clear discussion: key design difference of our work?}
These graph systems are either based on a vertex-centric model~\cite{malewicz2010pregel} or its variants (e.g., edge-centric~\cite{roy2013x}).
%where the exploration is from current active vertices or the \emph{frontier} to their neighbors, and the status of neighbors are determined by the messages sent by their parents. Those neighbors whose status have been changed become the new frontier. 
These models are in the strict BSP model~\cite{valiant1990bridging}. 
%that the frontier send messages in parallel and all messages are processed before the next step.
%
Different from them (and other asynchronous graph traversal efforts~\cite{han2017graphie, han2015giraph}), \Hammer uses delayed synchronization that is in the spirit of stale synchronization~\cite{ho2013more} where workers are running in an asynchronous fashion before synchronization, which makes it possible to retain high parallelism and also a low amount of distance computations. Moreover, as aforementioned in the implementation, due to the uniqueness of ANN, it is challenging to migrate many of these system designs to \Hammer directly.

\noindent{\bf Generic Search Schemes.}
Many efforts aim to parallelize various search schemes (e.g., BFS~\cite{shun2013ligra}, DFS~\cite{naumov2017parallel}, and Beam search~\cite{meister2020best}). Although \Hammer's parallel neighbor expansion design is inspired by prior parallel search algorithms on graphs, our work has a very different focus and aims to: 1) identify that ANN's convergence bottleneck comes from the fact that ANN requires to find many targets that may be (or not be) present in the graph---a search scenario that is very different from many previous graph search problems; 
% 2) discover that how the parallel neighbor expansion design can significantly reduce ANN convergence steps to near neighbors; 
2) several optimizations specifically tailored for reducing the number of distance computations and synchronization overhead from parallel neighbor expansion, such as staged search and redundant-expansion aware synchronization.

%% file: text/conclusion.tex
\section{Conclusion}\label{sec:conclusion}

%This work offers a parallel graph-based ANN search to reduce query latency.
This work looks into the problem of accelerating graph-based ANN search on multi-core systems, performing comprehensive studies to reveal multiple challenges and opportunities to exploit intra-query parallelism for speeding up ANN search. 
Based on the detailed performance characterization, we propose \Hammer, a similarity search algorithm that takes advantage of multi-core CPUs to significantly accelerate search speed without comprising search accuracy. \Hammer consists of a set of advanced parallel designs, including parallel neighbor expansion, staged search, redundant-expansion aware synchronization, loosely synchronized visit map, and cache friendly neighbor grouping, systematically addressing all the identified challenges. Evaluation results show that \Hammer outperforms two state-of-the-art methods NSG and HNSW by up to $37.7\times$ and $76.6\times$ on a wide range of real-world datasets ranging from million to billion data points. 

%Our future work will optimize the parallelization among different queries to provide good query throughput. Our ultimate goal is to let user choose a certain recall goal or latency goal, and then let \Hammer switch between different optimization techniques which aim on either intra-query parallelism or inter-query parallelism.